\documentclass[acmsmall]{acmart}

\AtBeginDocument{%
  \providecommand\BibTeX{{%
    \normalfont B\kern-0.5em{\scshape i\kern-0.25em b}\kern-0.8em\TeX}}}

\copyrightyear{2023}
\acmYear{2023}

\usepackage{multirow}
\usepackage{makecell}
\usepackage{pgf-pie}
\usepackage{tikz,stackengine}
\usepackage{caption}
\usepackage{subcaption}
\usepackage{mathrsfs}
\usepackage{titlesec}
\usepackage{enumitem}
\usepackage{bm} 

\titleformat{\subsubsection}
  {\normalfont\normalsize\bfseries}{\thesubsubsection}{1em}{}
\titlespacing*{\subsubsection}{0pt}{3.25ex plus 1ex minus .2ex}{1.5ex plus .2ex}

\usepackage{pgfplots}
\usepackage{pgfplotstable}

\usepackage{changepage} 

\usepackage[most]{tcolorbox}

\newtcbtheorem{Summary}{Summary of RQ\bfseries}{enhanced,drop shadow={black!50!white},
  coltitle=black,
  top=0.3in,
  attach boxed title to top left=
  {xshift=1.5em,yshift=-\tcboxedtitleheight/2},
  boxed title style={size=small,colback=gray}
}{summary}

\newtcolorbox[auto counter]{summary}[1][]{title={\bfseries Summary~\thetcbcounter},enhanced,drop shadow={black!50!white},
  coltitle=black,
  top=0.3in,
  attach boxed title to top left=
  {xshift=1.5em,yshift=-\tcboxedtitleheight/2},
  boxed title style={size=small,colback=gray},#1}

\begin{document}

\title{Unraveling Code Clone Dynamics in Deep Learning Frameworks}

\author{Maram Assi}
\email{maram.assi@queensu.ca}
\affiliation{
  \institution{Queen's University}
  \city{Kingston}
  \state{Ontario}
  \country{Canada}
}

\author{Safwat Hassan}
\email{safwat.hassan@utoronto.ca}
\affiliation{
  \institution{University of Toronto}
  \city{Toronto}
  \state{Ontario}
  \country{Canada}
}

\author{Ying Zou}
\email{ying.zou@queensu.ca}
\affiliation{
  \institution{Queen's University}
  \city{Kingston}
  \state{Ontario}
  \country{Canada}
}


\begin{abstract}
\textit{Deep Learning (DL)} frameworks play a critical role in advancing artificial intelligence, and their rapid growth underscores the need for a comprehensive understanding of software quality and maintainability. DL frameworks, like other systems, are prone to code clones. Code clones refer to identical or highly similar source code fragments within the same project or even across different projects. Code cloning can have positive and negative implications for software development, influencing maintenance, readability, and bug propagation. While the existing studies focus on studying clones in DL-based applications, to our knowledge, no work has been done investigating clones, their evolution and their impact on the maintenance of DL frameworks. In this paper, we aim to address the knowledge gap concerning the evolutionary dimension of code clones in DL frameworks and the extent of code reuse across these frameworks. We empirically analyze code clones in nine popular DL frameworks, i.e., \textit{TensorFlow}, \textit{Paddle}, \textit{PyTorch}, \textit{Aesara}, \textit{Ray}, \textit{MXNet}, \textit{Keras}, \textit{Jax} and \textit{BentoML}, to investigate (1) the characteristics of the long-term code cloning evolution over releases in each framework, (2) the short-term, i.e., within-release, code cloning patterns and their influence on the long-term trends, and (3) the file-level code clones within the DL frameworks. Our findings reveal that DL frameworks adopt four distinct cloning trends: \textit{"Serpentine"}, \textit{"Rise and Fall"}, \textit{"Decreasing"}, and \textit{"Stable"} and that these trends present some common and distinct characteristics. For instance, bug-fixing activities persistently happen in clones irrespective of the clone evolutionary trend but occur more in the \textit{"Serpentine"} trend. Moreover, the within-release level investigation demonstrates that short-term code cloning practices impact long-term cloning trends. The cross-framework code clone investigation reveals the presence of \textit{functional} and \textit{architectural adaptation} file-level cross-framework code clones across the nine studied frameworks. We provide insights that foster robust clone practices and collaborative maintenance in the development of DL frameworks.

\end{abstract}



\maketitle

\section{Introduction}
\label{sec:introduction}

Deep Learning (DL) frameworks are pivotal in shaping the future of artificial intelligence and machine learning. The Global DL Market is anticipated to grow a Compound Annual Growth Rate (CAGR) of 51.1\% from 2022 to 2030\footnote{https://www.acumenresearchandconsulting.com/deep-learning-market}. 
Moreover, empirical data highlights the substantial popularity of DL repositories on GitHub, providing further evidence of the significance of DL frameworks. For instance, the \textit{TensorFlow} framework has attracted remarkable attention, accumulating over 150,000 stars within an eight-year span \cite{GithubRanking}. Additionally, \textit{TensorFlow} attains an impressive count of over 88,000 forks, making it among the top 5 globally recognized repositories on GitHub. These metrics affirm the widespread acknowledgment of the pivotal role of these frameworks in the artificial intelligence software development landscape. 

Code cloning is the replication or duplication of source code fragments within a software project \cite{Roy_2007}. It can manifest as either exact or similar copies of code segments. While code cloning might alleviate developers' workload \cite{4023973}, multiple studies have demonstrated the potential negative impact of code clones on the development and maintenance of software systems \cite{4658071, 1610609}. Therefore, developers should be aware of the costs of the risk of code cloning. The evolution and impact of code clones are well studied in traditional systems \cite{Barbour_2013, Mondal_2012, 8090146, 7476632}.

DL code differs from traditional system code in terms of the development paradigm and coding practices \cite{8804457}. A recent study demonstrates that clone occurrences are higher in DL systems as compared to traditional code \cite{10.1007/s10664-021-10099-x}. It is not clear if the clone analysis of traditional software could be adapted to DL software. Despite the rising popularity of DL software, only two prior studies have explored code clones within the DL domain. Jebnoun et al. \cite{10.1007/s10664-021-10099-x} study clones in Python, C\#, and Java-based DL applications, shedding light on the prevalence of code clones during model creation, training, and data preprocessing. In a recent study, Mo et al. \cite{10.1145/3607181} emphasize co-changed clones within DL applications. While the existing studies focus on studying clones in DL-based applications, to our knowledge, no work has been done investigating clones, their evolution and their impact on the maintenance of DL frameworks.

To address the knowledge gap concerning the evolution of code cloning and its implications within the DL frameworks, in this work, we conduct an empirical study to gain a better understanding of the evolutionary dimension of code clones in DL frameworks and the extent of code reuse across these frameworks. More specifically, we analyze long-term code cloning trends over releases and short-term code cloning patterns within releases. Our goal is to offer insights on better managing and mitigating clones in DL frameworks by identifying the characteristics of short-term patterns and their impact on long-term trends. Additionally, we conduct a cross-framework clone study to provide insights into the code commonalities and collaborative work across DL frameworks.

We conduct experiments on nine popular DL frameworks, i.e., \textit{TensorFlow}\footnote{https://github.com/tensorflow/tensorflow.git}, \textit{Paddle}\footnote{https://github.com/PaddlePaddle/Paddle.git}, \textit{PyTorch}\footnote{https://github.com/pytorch/pytorch.git}, \textit{Aesara}\footnote{https://github.com/aesara-devs/aesara.git}, \textit{Ray}\footnote{https://github.com/ray-project/ray.git}, \textit{MXNet}\footnote{https://github.com/apache/mxnet.git}, \textit{Keras}\footnote{https://github.com/keras-team/keras.git}, \textit{Jax}\footnote{https://github.com/google/jax.git} and \textit{BentoML}\footnote{https://github.com/bentoml/BentoML.git}. \noindent Our empirical investigation yielded the following findings answering the studied research questions:\\

\noindent\textbf{RQ1: What are the characteristics of the long-term trends observed in the evolution of}
\begin{adjustwidth}{0.8cm}{}
     \textbf{code clones within DL frameworks over releases?}\\ 
       Our goal is to explore the characteristics of the long-term trends of the evolution of code clones over releases within DL frameworks.
       We observe that the four cloning trends, i.e., \textit{"Serpentine"}, \textit{"Rise and Fall"}, \textit{"Decreasing"} and \textit{"Stable"} exhibit distinct and common characteristics. The reduction in cloned code size in the \textit{"Rise and Fall"} and \textit{"Decreasing"} trends is attributed to factors, such as code refactoring, the reuse of third-party libraries, and the removal of code clones linked to feature elimination. Furthermore, our findings indicate that bug fixing is a consistent activity persisting throughout framework lifespans among all the trends but has a higher presence in the frameworks of the \textit{"Serpentine"} trend.\\
 \end{adjustwidth}

\noindent\textbf{RQ2: What are the characteristics of within-release code cloning patterns and do these}
\begin{adjustwidth}{0.8cm}{}
     \textbf{patterns contribute to the overarching long-term trends in code cloning?}\\ 
     Our goal is to identify the characteristics of short-term code cloning patterns within-release and explore their potential impact on the long-term trends in code cloning. Our results demonstrate that within-release code cloning patterns, i.e., \textit{"Ascending"}, \textit{"Descending"}, and \textit{"Steady"} patterns, impact long-term code cloning trends. In addition, we unravel the characteristics of the within-release patterns. For instance, \textit{"Ascending"} within-release code cloning pattern is associated with decreased committer involvement in clone pairs, suggesting that fewer committers may lead to a higher likelihood of an increase in code cloned size.\\
 \end{adjustwidth}
     
\noindent\textbf{RQ3: How do code clones manifest and evolve across different DL frameworks?}
\begin{adjustwidth}{0.8cm}{}
We aim to conduct a cross-framework clone detection to identify and analyze similarities in code across different DL frameworks. We find that cross-framework file-level code clones exist within DL frameworks, and they fall into two categories: \textit{functional} and \textit{architectural adaptation} code clones. Cross-framework code clones undergo a gradual disappearance, attributed to functionality evolution, code divergence, function deprecation and framework restructuring.\\
 \end{adjustwidth}

The main contributions of our work are as follows:
\begin{enumerate}
    \item We analyze the characteristics of the evolution of code clones in DL frameworks throughout their entire lifespan. The analysis of distinct cloning trends provides insights to enhance the efficiency and maintainability of DL frameworks over time.
    \item We offer insights into code cloning development patterns within releases that influence code clone evolution. We identify factors affecting cloned code size and projecting long-term clone trends and provide insights into well-maintained frameworks within the DL framework community.
    \item We conduct a cross-framework code clone analysis within DL frameworks. Our investigation of the code commonality and clones across the frameworks can potentially motivate collaborative efforts within the DL community.
    \item We provide a replication package\footnote{https://github.com/mia1q/code-clone-DL-frameworks/} for our approach, including the list of DL frameworks, the genealogy results obtained, and the scripts necessary to reproduce our study.
\end{enumerate}

\textbf{Paper organization.} The remaining part of the paper is organized as follows. Section~\ref{sec:experimental_setup} presents the experimental setup. Section~\ref{sec:experimental_results} illustrates the motivation, approach, and findings of our research questions. Section~\ref{sec:discussion} discusses the implications of our work. Section~\ref{sec:threats} describes the potential threats of this study. Section~\ref{sec:related_work} discusses the related work. Finally, Section~\ref{sec:conclusion} concludes the study and discusses future work.

\section{Experiment Setup}
\label{sec:experimental_setup}
\begin{figure*}
\centering
\includegraphics[width=1\linewidth]{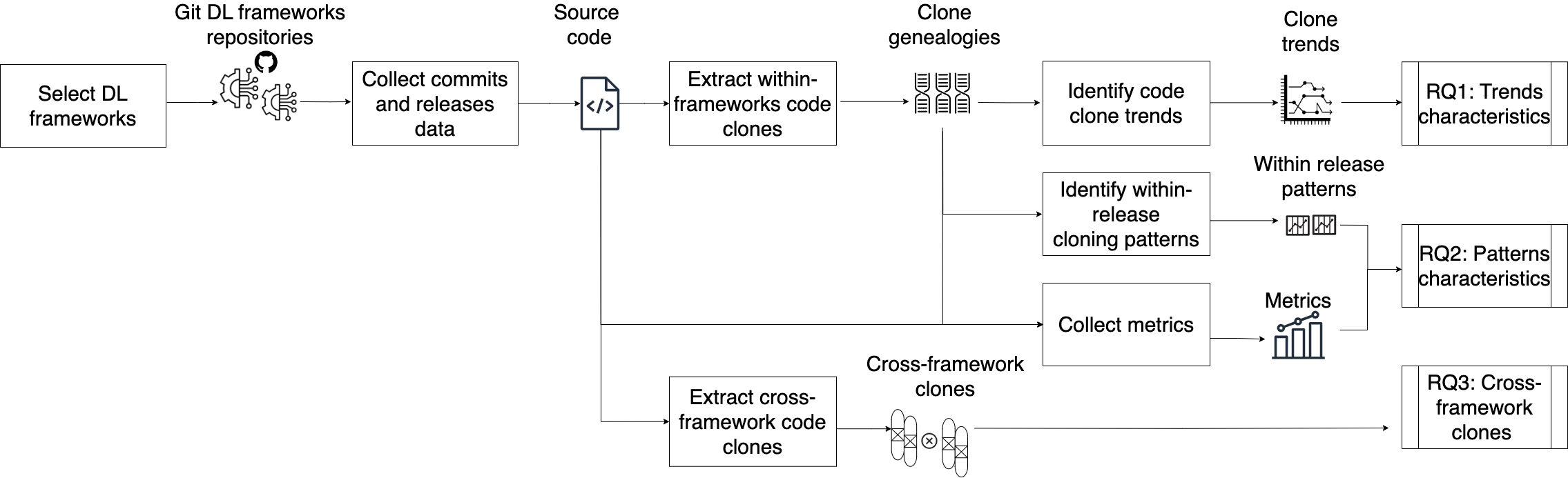}
\caption{An overview of our approach for analyzing code cloning in DL frameworks.}
\label{fig:approach_overview}
\end{figure*}

In this section, we present the experimental setup. Figure~\ref{fig:approach_overview} depicts the overview of our approach for analyzing code cloning in DL frameworks. In the first step, we select the DL frameworks. Then, we collect the commit and release data to construct the source code version history for each framework. Next, we extract the code clones within individual frameworks to build the clone genealogies and collect relevant metrics. In the final step, we extract the code clones across all the DL frameworks to answer RQ3.

\subsection{Deep Learning frameworks selection}
\label{dataset}
DL frameworks provide a robust foundation for implementing DL applications, driving significant technological advancements such as imaging and autonomous vehicles. To identify the target DL frameworks for our study, we follow  Du et al. \cite{9716780} approach and refer to the widely studied frameworks in the recent existing work \cite{10.1145/3338906.3338955, 10.1007/s10664-021-10099-x, 10.1145/3368089.3409761, 10.1145/3587155, Humbatova_202, 8355444, 10.1145/3514094.3534167, 9463133, Wang2019}. We investigate 14 candidates, including \textit{TensorFlow}, \textit{Chainer}, and \textit{Keras}. Python has emerged as the most widely adopted programming language for DL applications \cite{8595219}. Therefore, we tailor our data collection approach to include only Python frameworks. To ensure the robustness of our dataset, we manually check the GitHub repositories of the DL frameworks and read the framework readme file to ensure it is DL-specific. We end up with the following nine DL frameworks: \textit{TensorFlow}, \textit{Paddle}, \textit{PyTorch}, \textit{Aesara}, \textit{Ray}, \textit{MXNet}, \textit{Keras}, \textit{Jax} and \textit{BentoML}.

\subsection{Commit and releases data collection}
\label{data_collection}

We extract the historical data of each repository to study the evolution of code clones and identify the evolution trends. In particular, we collect the commit history to extract the corresponding source code over time and the release information to be able to observe the trends in code cloning across different releases.\\
 
\noindent\textbf{Commit collection and source code extraction.} For every framework, we obtain all commits by utilizing the \texttt{git log -- pretty=format:"\%h,\%ae, \%ai, \%s"} command, which includes commit information, such as the commit log message, author and committer details, and commit dates. For every commit, we compute the snapshot size in terms of Python source lines of code (SLOC) using the Cloc library \footnote{https://github.com/AlDanial/cloc} version \textit{v-1.96}.\\

\noindent\textbf{Release collection and pre-processing.} We gather the framework releases through the GitHub API, including their version numbers, release dates, and the associated release notes. A version string is composed of three numerical components delimited by periods, e.g., “1.7.2”. We follow the existing work \cite{Yang2022AnES} and adopt semantic versioning (i.e., version strings represented by three numbers separated by dots) [39] to categorize the releases in our study into three main types: \textit{major}, \textit{minor}, and \textit{patch}. \textit{Major} releases are associated with software's API changes, \textit{minor} releases to the addition of new features and \textit{patch} to backward-compatible bug fixing. A major release corresponds to a modification in the first number, a minor release to a modification in the second number, and a patch release occurs in the third number. To determine the release type, we compare the version strings of the current release \(R_{i}\) and the preceding release \(R_{i-1}\). If there are changes in multiple numbers, the precedence follows major over minor or patch, and minor over patch. In our study, we exclude \textit{patch} releases and include only \textit{major}, \textit{minor} as they represent stable releases. Table~\ref{tab:dataset_statistics} summarizes the statistics of the collected DL frameworks. \\

\begin{table}[h]
    \centering
    \caption{The Descriptive Statistics of the Dataset.}            
    \label{tab:dataset_statistics}
    \begin{tabular}{lccc}
        \toprule
        \textbf{Framework} & \textbf{First commit} & \textbf{\# of commits} & \textbf{\# releases}\\
        \midrule      
        Aesara & 2008-01-06 & 29,600 & 14\\
        Keras & 2015-03-27 & 7,787 & 9\\
        MXNet & 2015-04-30 & 11,865 & 20\\
        Ray & 2016-02-07 & 16,403 & 19\\
        PyTorch & 2016-05-02 & 48,561 & 18\\
        Paddle & 2016-08-29 & 36,531 & 22\\
        TensorFlow & 2018-03-04 & 131,854 & 46\\
        Jax & 2018-11-17 & 13,868 & 7\\
        BentoML & 2019-04-01 & 2,100 & 11\\
\bottomrule

    \end{tabular}
\end{table}

\subsection{Within-framework code clone extraction}
\label{sec:clone_detection}
To identify code clones within one framework, we employ NiCad tool \cite{4556129}, a well-known and most-used text-based clone detection tool \cite{zakerinasrabadi2023systematic}, which exhibits strong capabilities to detect both exact and near-miss clones at the block and function level with high precision and recall \cite{4976382}. We use the latest available version \textit{NiCad6.2} of NiCad \footnote{https://www.txl.ca/txl-nicaddownload.html}. We follow the standard configuration of NiCad, i.e., dissimilarity threshold of 30\% and minimum cloned code size of 5 lines, used in recent related work \cite{10.1007/s10664-021-10099-x, 10.1145/3607181} as with these settings, NiCad is reported to be very accurate in clone detection. We detect three code clone types: Type 1, Type 2, or Type 3, and we choose clones at the functions level.\\

\noindent\textbf{Snapshot code clone detection.}
We apply the following steps to every commit of the studied DL framework to detect clones. Firstly, we use the \texttt{git checkout} command to extract the framework snapshot corresponding to a specific commit from GitHub. Secondly, since in our study, we focus on code clones within production code, we follow the existing work \cite{barbour2018investigation} to exclude test-related code by identifying files and folders with the "test" keyword. Lastly, we use NiCad to detect the code clones. The output of the clone detection is pairs of functions, where two functions that are cloned, i.e., highly similar to each other, constitute a clone pair.\\

\noindent\textbf{Clone genealogy generation.} Clone pairs can undergo code changes throughout the development and maintenance stages of a framework. As a result of a change, a clone pair can maintain a \textit{consistent} state, i.e.,  cloned status, or diverge to an \textit{inconsistent} state. A genealogy represents the historical evolution of a pair of code clones. To generate the clone genealogy of each clone pair, we track its modification in every commit along the framework commit list. Similar to the existing work \cite{10.1007/s10664-023-10292-0, barbour2018investigation}, we conduct the below steps to generate the genealogy of every clone pair:

\begin{enumerate}
\renewcommand\labelenumi{\arabic{enumi}.} 
\item For a given commit \(C_{i}\), we identify the files changed. If a changed file changes a clone pair, we proceed by comparing \(C_{i}\) to the previous commit \(C_{i-1}\) to verify if the change was made to the clone pair. We use the \texttt{diff} command and a python third party library \textit{whatthepatch}\footnote{https://pypi.org/project/whatthepatch/} to map the lines corresponding to the beginning and end of the function in both commits \(C_{i}\) and \(C_{i-1}\). To handle the case where a file name was modified, we map the renamed file to the original file by performing the following steps as introduced in the existing work \cite{10.1007/s10664-023-10292-0, barbour2018investigation}. First, for each commit, we extract the pairs of newly added and deleted files between the current and preceding commit. Second, we compare the code similarity and consider that a file was renamed if the content of the new file is similar to the content of the old file. We use the below git command to construct the pairs of renamed files between two commits: \texttt{git diff [old-commit] [new-commit] --name-status -M}

\item We check if a code change, i.e., code lines addition or deletion, is made in \(C_{i}\) within the boundaries of the clone pair. If so, we verify if this clone pair exists in the clone snapshot corresponding to commit \(C_{i+1}\). If the clone pair exists in \(C_{i+1}\), we attribute the state \textit{consistent} to it; otherwise, we consider the pair state as \textit{inconsistent}.

\item We reiterate this process, i.e., steps 1 and 2, for every commit in the framework repository or until at least one of the clones is deleted.

\end{enumerate}

\subsection{Cross-framework file-level code clone extraction}
\label{sec:cross_projet_clone}
To identify code clones across several DL frameworks, we employ SourcererCC \cite{10.1145/2884781.2884877}, a token-based clone detection tool capable of identifying both exact and near-miss clones within extensive inter-framework repositories. We select SourcererCC as it outperforms other clone detectors, e.g., Nicad, for large-scale cross-projects clone detection \cite{DBLP:journals/corr/abs-1808-00106, 9679816} and it was used by recent cross-project clone related work \cite{8530022, 9679816, 9047640} for its high performance. We use the publicly available version\footnote{https://github.com/Mondego/SourcererCC} on GitHub. SourcererCC performs clone detection in two steps. Firstly, it employs a tokenizer where the programming language parameters, such as file extension, are set. In our case, we set the file extension to ".py". Secondly, the actual clone detection is executed. For this step, the similarity thresholds parameter is specified. Existing work used different values for the similarity threshold. For example, Rahman et al. \cite{9047640} used 70\% whereas Chochlov et al. \cite{9978241} used 50\%. In our study, we vary the value between 60\% and 80\%. We adopt the 60\% similarity threshold in this paper as it detects meaningful code clones across frameworks.

\subsection{Metrics collection}
\label{sec:metric_calculation}

After we generate the code clone snapshots and genealogies, we collect cloned metrics for every framework. The collected clone metrics capture information about the snapshot and the genealogy of a clone pair to investigate the relationship between the collected metrics and the cloned code size. In total, we collect 29 metrics, representing (1) product metrics, (2) development community metrics, (3) genealogy metrics and (4) code metrics. These metrics were used in existing work \cite{barbour2018investigation, 10.1007/s10664-023-10292-0, 10.1007/s10664-018-9645-2}. We describe below each of the categories and present the collected metrics in Table \ref{tab:metrics_collection}.\\

\noindent\textbf{Product metrics.} Product metrics are the attributes related to the snapshot, i.e., commit. The product metrics are collected from the snapshot of the framework that contains the clone pair. Example metrics include the number of siblings a clone has in the version where it is introduced, the size of the clone and the number of clone groups.\\

\noindent\textbf{Development community metrics.} Development community metrics include metrics representing the involvement of contributors in the development of clone pairs. For example, the number of committers, i.e.,  the person who applied the commit to the repository, and authors, i.e., the person who originally wrote the changes introduced in the commit, associated with the genealogy of clone pairs that provide insights into the collaborative efforts, i.e., number of individuals that have applied and integrated changes related to the clone pairs, and individual contributions to the codebase over time.\\

\noindent\textbf{Genealogy metrics}. Genealogy metrics capture the state changes in the history of clone pairs and the history of changes in a clone pair. An example of a genealogy metric is the ratio of inconsistent changes in the whole framework code clone genealogy.\\

\noindent\textbf{Code metrics}. Code metrics include metrics related to the code characteristics of the method that contains a clone, such as cyclomatic complexity and the number of declaration and execution statements. To analyze code metrics, we use the Understand tool\footnote{https://scitools.com/}, a reverse engineering tool capable of analyzing product metrics from the source code of software systems. First, we run the Understand tool on all the snapshots of DL frameworks and extract the metrics on the function level. We obtain metrics such as the number of declarative statements in a function and the complexity of a function. Second, we map the functions obtained by the Understand tool to the clone code to compute the metrics for the cloned metrics.\\

\begin{table}[h]
    \centering
    \caption{Collected code clone metrics.}
    \label{tab:metrics_collection}
    \small
    \begin{tabular}{ll}
        \toprule
        \textbf{Metrics} & \textbf{Description}\\ \midrule
        \multicolumn{2}{c}{\textbf{Product metrics}}\\ \midrule
        \textit{NumCP} & The total number of clone pairs at any specific commit.\\        
        \textit{NumCPGroups} & The unique number of groups at any specific commit.\\
        \textit{MedianCLOC} & The median number of cloned lines of code \cite{barbour2018investigation}.\\
        \textit{MaxCLOC} & The maximum number of cloned lines of code \cite{barbour2018investigation}.\\
        \textit{MedianCSib} & The median number of siblings of the clone pairs at any specific commit \cite{10.1007/s10664-023-10292-0}.\\
         \textit{MaxCSib} & The maximum number of siblings of the clone pairs at any specific commit \cite{10.1007/s10664-023-10292-0}.\\
        \textit{CPMedianAgeDays} & The median clone pair age in terms of the number of days \cite{10.1007/s10664-023-10292-0}. \\
        \textit{CPMedianAgeCommits} & The median clone pair age in terms of the number of commits \cite{10.1007/s10664-023-10292-0}. \\
        \textit{NumUniqueFiles} & The unique number of files that contain clone pairs.\\
        \textit{NumAbstClasses} & The number of abstract classes containing code clones.\\
        \textit{MedCodeSim} & The median of code clone similarity.\\
        \textit{RatioSync} & The ratio of clone pairs that have a final “consistent” change in the genealogy \cite{barbour2018investigation}.\\
        \textit{RatioLatePropCons} & The ratio of clone pairs having late propagation, which end in a consistent \\
        & change \cite{barbour2018investigation}.\\
        \textit{RatioLatePropIncons} & The ratio of clone pairs having late propagation, which end in an inconsistent\\
        & change \cite{barbour2018investigation}.\\  
        \midrule

        \multicolumn{2}{c}{\textbf{Development community metrics}} \\ \midrule 
       \textit{NumCommittorsGen} & The number of committers involved in the genealogy \cite{10.1007/s10664-023-10292-0}.\\
        \textit{NumAuthorsGen} & The number of authors involved in the genealogy\\
         \textit{MedianNumComChanges} & The median number of commits in the genealogy by a specific committer \cite{10.1007/s10664-023-10292-0}.\\
        \textit{MedianNumAutChanges} & The median number of commits in the genealogy by a specific author.\\
        \midrule

        \multicolumn{2}{c}{\textbf{Genealogy metrics}} \\ \midrule 
        \textit{NumCPG} & The number of clone pairs in the genealogy.\\ 
        \textit{NumChanges} & The total number of changes in the genealogy.\\
        \textit{NumBursts} & The number of change bursts on a clone. A change burst is a consecutive \\ & change with a maximum distance of one day between the changes \cite{barbour2018investigation}.\\
        \textit{NumInconsChanges} & The number of inconsistent changes within the genealogy \cite{barbour2018investigation}.\\
        \textit{NumDiverChanges} & The number of divergent changes (pattern "CI") in the genealogy \cite{barbour2018investigation}.\\
        \textit{NumResyncChanges} & The number of resynched changes (pattern "IC") in the genealogy \cite{barbour2018investigation}.\\
        \midrule
        
        \multicolumn{2}{c}{\textbf{Code metrics}} \\  \midrule 
        \textit{MedianCompl} & The median cyclomatic complexity of the cloned code \cite{10.1007/s10664-018-9645-2}.\\
        \textit{MedianNumDeclStmt} & The median number of declaration statements in a cloned code.\\
        \textit{MedianNumExecStmt} & The median number of execution statements in a cloned code. \\
        \textit{MedianRatioCommentCode} & The median ratio comment to code in cloned code \cite{10.1007/s10664-018-9645-2}. \\
        \textit{MedianSumComplAbstMod} & The median sum of cyclomatic complexity of all the classes containing \\
        & cloned functions. \\
        \bottomrule
    \end{tabular}
\end{table}

\subsection{Lifelong code cloning trends identification from DL frameworks }
\label{clone_evolution_trends}
Our goal is to discern the lifelong evolution, i.e., long-term, trends of code clones within DL frameworks to gain insights into the historical and chronological changes in cloned code over multiple releases. This analysis can provide valuable information about the stability, growth, or reduction of code clones over time, aiding in the assessment of software evolution and maintenance practices within DL frameworks. For every DL framework, we build a time series, i.e., a temporal representation, capturing the historical and chronological changes to the cloned code over the releases of a framework. Then, we group similar timer series patterns together and identify the code clone trends over the releases. We elaborate on our approach in the following steps.\\

\noindent\textbf{Step 1: Building time series data.} To build the time series that represents the historical evolution of code clones, we start by calculating the clone coverage. Clone coverage represents the proportion of cloned code in relation to the overall codebase \cite{Inoue2021}. The clone coverage is calculated as follows:

\begin{align*}
\text{clone\_coverage}_i & = \frac{\text{clone\_size}_i}{\text{SLOC}_i}
\end{align*}
\noindent where \(clone\_coverage_{i}\) represents the clone coverage corresponding for \textit{commit i}. The \(clone\_size_{i}\) represents the total number of lines of code belonging to clones at \textit{commit i}. The \(SLOC_{i}\) represents the lines of source code of the overall repository at \textit{commit i}. 

Our goal is to construct a release-level clone coverage time series for every framework. We choose the release-level because it represents a stable granularity as opposed to the commit-level granularity that would have been a fine granularity, given the small and/or temporary modifications that a commit can introduce
\cite{10.1145/3127005.3127008}. We calculate the clone coverage for every revision, i.e., commit, using the code clone data extracted through the NiCad tool detailed in section \ref{sec:clone_detection} and the SLOC of the framework snapshot at the commit. Therefore, for every release, we compute the median code clone coverage for all the commits within the time interval between two consecutive releases. We refer to the list of releases extracted from GitHub as described in section \ref{data_collection} and the associated release dates to isolate the commits falling within the release date intervals. Then, we calculate the median release clone coverage as follows:

\begin{align*}
\text{clone\_coverage}_{R_{j}} = \text{median clone coverage of all commits in the release}  {R_{j}}
\end{align*}

By the end of this step, we obtain the median value of code clone coverage for every release of every framework. For each DL framework, we construct a time series representing the evolutionary trajectory of clone coverage across multiple releases. By the end of this step, we obtain nine unique time series, one for each framework.\\

\noindent\textbf{Step 2: Clustering time series.} We utilize time series clustering to group time series presenting similar patterns in code cloning. This approach involves the selection of (i) a method for measuring distances, (ii) a clustering algorithm, and (iii) an optimal number of clusters, as we elaborate on below.

\begin{itemize}[label=--]
  \item \textit{Dynamic Time Warping (DTW).} DTW is used as a distance measurement method \cite{10.5555/3000850.3000887} to align and measure the similarity between two time series that may have different temporal characteristics. In our context, it is used to compare and align the time series representing code clone coverage across multiple releases of DL frameworks. Different DL frameworks may have a distinct number of software releases, resulting in time series of varying lengths. To measure the similarity, i.e., distance, between time series data of different lengths, we adopt the DTW.\\

\begin{figure*}
\centering
\includegraphics[width=1\linewidth]{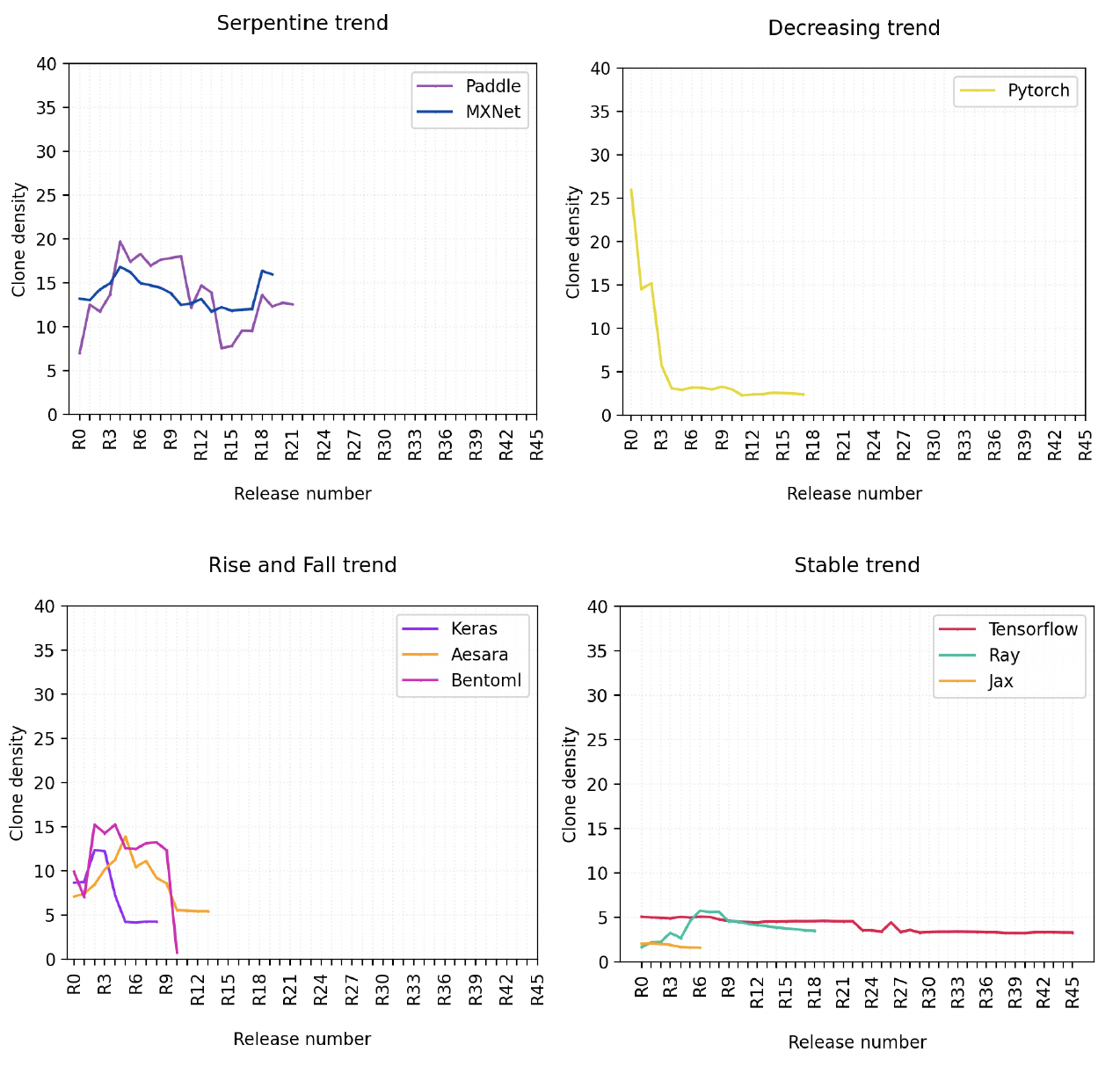}
\caption{The four code clones trends exhibited by DL frameworks.}
\label{fig:clone_evolution_trends}
\end{figure*}

    \item \textit{TimeSeriesKMeans clustering algorithm} is used for time series clustering \cite{WARRENLIAO20051857} to identify patterns in the evolution of the nine DL frameworks. K-means clustering is a widely used unsupervised machine learning technique that is particularly effective for partitioning data into distinct groups or clusters based on the similarity of the data points. K-Means treats each time series as a point in a multidimensional space, aiming to group similar time series together in clusters. In our analysis, we employ tslearn \footnote{https://tslearn.readthedocs.io/en/stable/gen\_modules/tslearn.clustering.html\#module-tslearn.clustering} implementation of K-means for time series, TimeSeriesKMeans \footnote{https://tslearn.readthedocs.io/en/stable/gen\_modules/clustering/tslearn.clustering.TimeSeriesKMeans.html} to identify and categorize the trends present in the time series data representing code clone coverage within DL frameworks.\\

    \item \textit{Optimal number of clusters}. Following prior work \cite{10315108}, we employ the silhouette score\cite{ROUSSEEUW198753} to identify the most suitable number of clusters. The silhouette score assesses clustering quality by measuring how close data points are to their cluster compared to other clusters. The silhouette score ranges from -1 to 1, where a score close to 1 suggests well-defined clusters, and a score above 0.5 indicates a good clustering configuration. It leverages the output of the chosen clustering algorithm, e.g., K-Means, to determine the optimal number of clusters, aiming for the highest average silhouette score, which indicates well-separated and coherent clusters. We employ the silhouette score function\footnote{https://tslearn.readthedocs.io/en/latest/gen\_modules/clustering/tslearn.clustering.silhouette\_score.html} from tslearn. We use a range of numbers of clusters, k, between 2 and 8. As a result, we obtain 4 optimal numbers of clusters with a silhouette score of 0.63, suggesting a good quality of clusters. Additionally, we rely on human judgement that aligns with the quantitative metrics.\\
  
\end{itemize}

\noindent\textbf{Our approach identifies four trends, i.e., \textit{"Serpentine"}, \textit{"Rise and Fall"}, \textit{"Decreasing"}, and \textit{"Stable"}, in the evolution of clones observed in DL frameworks} as illustrated in Figure \ref{fig:clone_evolution_trends}. We describe the features of every pattern below:
\begin{itemize}[label=--]
 \item The \textit{"Serpentine"} trend reflects a cyclic, oscillatory pattern in code clones, where coverage alternates between highs and lows, presenting fluctuations in code clone coverage over successive releases. 
 \item The \textit{"Rise and Fall"} trend marks intermittent rises in clone coverage followed by an overall prolonged decrease, resulting in a downward trajectory over time. 

\item The \textit{"Decreasing"} trend represents a consistent and continuous reduction in clone coverage across successive releases.

\item The \textit{"Stable"} trend exhibits a relatively constant and consistent clone coverage with minimal fluctuations over releases. 
\end{itemize}

For example, while \textit{Paddle} and \textit{MXNet} display a \textit{"Serpentine"} trajectory, \textit{BentoML}, \textit{Keras}, and \textit{Aesara} follow a \textit{"Rise and Fall"} trajectory that shows a prolonged decrease despite intermittent rises. \textit{PyTorch} exhibits a consistent decreasing clone coverage, i.e., starting at 27\% at the first release and dropping to less than 3\% after 20 releases. \textit{TensorFlow}, \textit{Jax}, and \textit{Ray}'s clone coverage remain relatively stable, with minimal fluctuations observed across all releases remaining within a narrow range (less than 5\%). \\

\subsection{Within-release development patterns identification}
\label{within-release-patterns}

Our goal is to identify the within-release development patterns, i.e., the patterns of cloned code size evolution from one commit to another within a framework release. This analysis investigates how code clones evolve within each individual release and can help uncover the impact of the within-release patterns on the long-term code cloning trends. Our approach involves two key steps:\\

\noindent\textbf{Step 1: Building within-release cloned code size time series.} The evolution of code cloned code size between the commits of a release can be interpreted as time series data. To build the time series for every release of every framework, we select the commits that belong to every release interval by referring to the release date. Then, we calculate the cloned code size using the code clone data extracted through the NiCad tool detailed in section 2.2 for every commit. In total, we obtain 153 time series corresponding to the nine DL frameworks' releases.\\

\begin{figure*}
\centering
\includegraphics[width=1\linewidth]{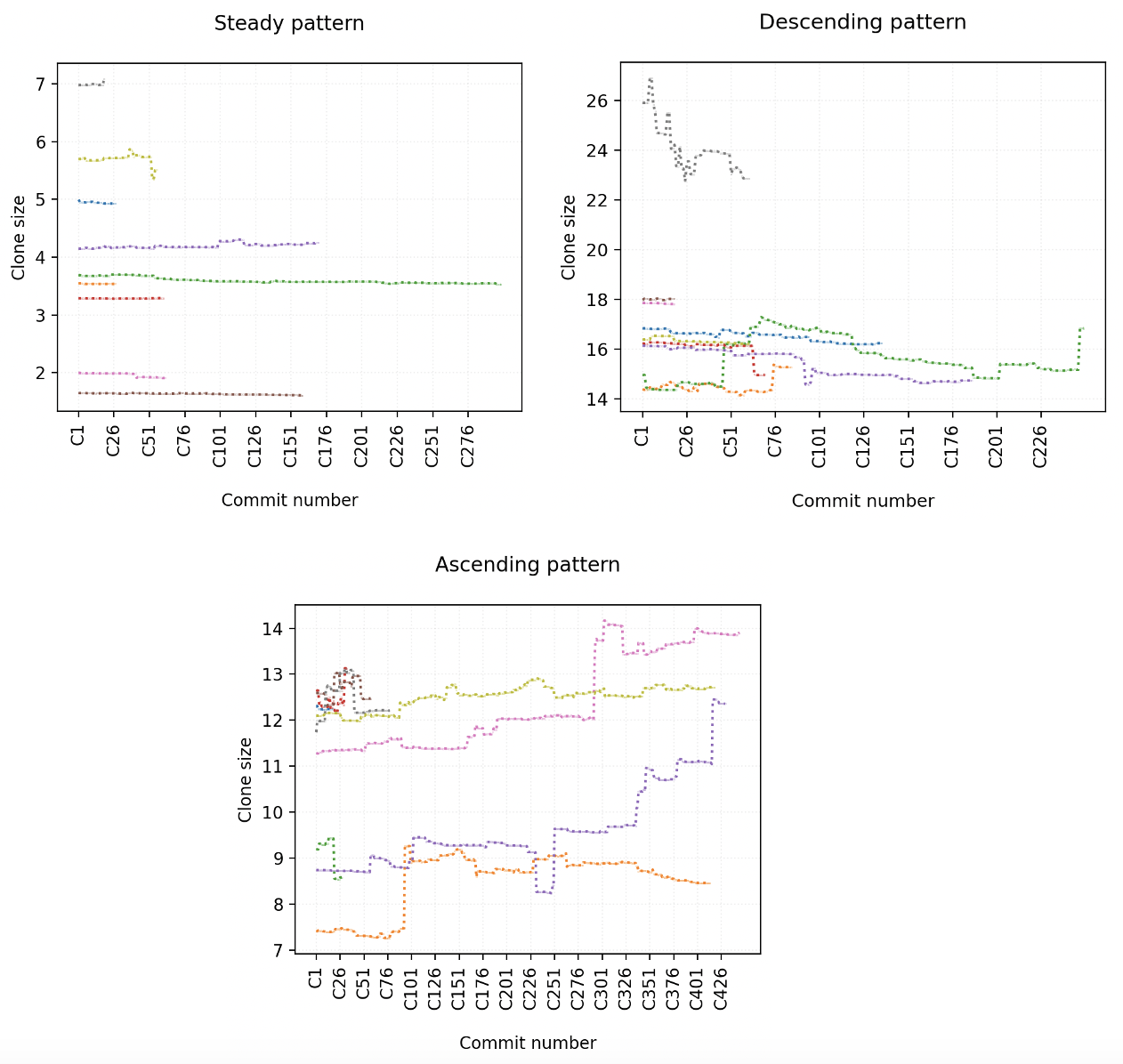}
\caption{Subset of the within-release time series \textit{"Steady"}, \textit{"Descending"} and \textit{"Ascending patterns"}. Every time series represents the evolution of clone size within a particular release of a DL framework.}
\label{fig:within_release_patterns}
\end{figure*}

\noindent\textbf{Step 2: Clustering within-release code clone time series.} Similarly to the clustering step in section \ref{clone_evolution_trends}, we leverage DTW and TimeSeriesKMeans to cluster the obtained time series. We determine the number of clusters using the silhouette score, and we obtain 3 as the optimal number of clusters, giving a silhouette score of 0.66, indicating a high quality of clusters. Figure \ref{fig:within_release_patterns} shows a subset of the time series clustered into the three \textit{"Ascending"}, \textit{"Descending"}, and \textit{"Steady"} patterns.\\

\noindent\textbf{The analysis of release-level time series reveals three distinct patterns, i.e., \textit{"Ascending"}, \textit{"Descending"}, and \textit{"Steady"}}. As depicted in Figure \ref{fig:within_release_patterns}, the \textit{"Ascending"} pattern signifies a consistent increase within a release, the \textit{"Descending"} pattern reflects a continuous decrease within a release, and the \textit{"Steady"} pattern suggests a relatively constant behaviour within a release. Table \ref{tab:time_patter_release_dist} shows the distribution of time series among these patterns. \\

\begin{table}
\centering
\caption{The distribution of release-level time series across the three within-release code cloning patterns.}
\label{tab:time_patter_release_dist}
\small
\begin{tabular}{lc*{11}c}
\toprule
      \textbf{Pattern}&  \textbf{\footnotesize TensorFlow} & \textbf{\footnotesize Paddle} & \textbf{\footnotesize PyTorch} & \textbf{\footnotesize Aesara} & \textbf{\footnotesize Ray} & \textbf{\footnotesize MXNet} & \textbf{\footnotesize Keras} & \textbf{\footnotesize Jax} & \textbf{\footnotesize BentoML} & \textbf{\footnotesize Total}\\ \midrule
Ascending  & 0  & 11 & 0  & 8 & 0  & 12 & 4 & 0 & 4 & 39\\
Descending & 0  & 8  & 4  & 0 & 0  & 9  & 0 & 0 & 4 & 25\\
Steady     & 38 & 2  & 14 & 4 & 19 & 0  & 4 & 5 & 3 & 89\\
\bottomrule
\end{tabular}
\end{table}

\section{Results and Analysis}
\label{sec:experimental_results}
In this section, we identify the different code clone evolution trends in DL frameworks over multiple releases and investigate the within-release development practices shaping the release-level clone trends. In addition, we study the cross-framework clones within DL frameworks. Precisely, we discuss motivation, approach and findings for the following research questions.

\subsection{RQ1: What are the characteristics of the long-term trends observed in the evolution of code clones within DL frameworks over releases?}

\label{sec:RQ1}
\subsubsection{Motivation}
The field of DL is marked by rapid growth and continual advancements. As DL frameworks rapidly evolve, they are prone to technical debt \cite{8812912, 10.1145/3379597.3387479}. Recent work \cite{10.1145/3379597.3387479, 10.1145/3607181} demonstrates that code clones are prevalent in DL applications. Mo et al. \cite{10.1145/3607181} find that the choice of underlying DL frameworks influences the frequency of code clones within DL applications. Jebnoun et al. \cite{10.1145/3379597.3387479} discover that code clones within DL systems exhibit a higher susceptibility to bugs compared to non-cloned code. In this RQ, we aim to investigate the characteristics (e.g.,  bug-proneness and community size) of the four identified lifelong code clone evolution trends and uncover any implications that they may have. We aim to enhance our understanding of code quality and maintenance over time by studying these characteristics.

\subsubsection{Approach} 
\label{rq1_approach}

Our goal is to explore the characteristics associated with the identified code clone trends. We study the characteristics of the trends along three dimensions: cloned code size trend, bug-proneness and community size.\\

\noindent\textbf{Investigating the decrease of clone coverage.} To better understand the rationale behind the decrease in clone coverage in the long-term descending trends, i.e., trends that exhibit a lower long-term clone coverage, we conduct a manual analysis. First, we investigate the evolution of the cloned code size, i.e., the number of lines of cloned code, associated with the median data point of each release to check if the decrease in clone coverage is associated with a decrease in the cloned code size. Then, for each release, we manually read the release notes and the commit messages of the commits associated with the release interval, thereby gaining insights into the rationale of the evolution of clone coverage over time.\\

\noindent\textbf{Investigating bug-proneness within trends.} We follow a two-step approach to study the bug-proneness of code clones. First, we apply the keyword-based heuristic introduced by Mockus et al. \cite{883028} and adopted in related work \cite{8919065, 10.1007/s10664-021-10099-x, 10.1145/3607181, 6080794} to identify bug-fixing commits. We consider a commit as a bug-fixing commit if the commit message contains any of the keywords \textit{bug, fix, wrong, error, fail, problem, and patch}. Second, we match the bug-fixing commits to the cloned code. Specifically, we employ the \texttt{git diff} command to retrieve the altered lines, i.e., added or deleted, within each bug-fixing commit and compare them to the clones identified by NiCad. If the modified lines for fixing a bug occurred within a clone code, we classify that clone as potentially susceptible to bugs. This approach allows us to pinpoint the bug-fixing commits occurring in clones.
After identifying the bug-fixing commits belonging to the cloned code, we plot the evolution of the bug-proneness of cloned code over releases. To quantify bug-proneness, we compute \(\text{bug-fix density}_i\), i.e., the ratio of bug-fixing commits in the cloned code introduced within a release interval by the total number of all commits to code clones introduced within a given release \textit{i}. This ratio is calculated as follows:

\begin{align*}
\text{bug-fix density}_i & = \frac{\delta \text{ (clone\_bugs\_fix\_commits}_i)}{\delta \text{ (all\_clone\_commits}_i)}
\end{align*}

\noindent where \(\delta(\text{clone\textunderscore bugs \textunderscore fix \textunderscore commits}_i)\)
 represents the commits introduced within release \textit{i} interval and classified as bug-fixing and \(\delta(\text{all\textunderscore clone \textunderscore commits}_i)\) represents the total number of commits that occurred in cloned code within the release \(\text{}_i\). A bug-fix density ratio of 0 signifies no bug-fixing commits in the cloned code during a release, while a ratio of 1 indicates that all commits within that release for the cloned code are bug-fixing commits. By tracking the evolution of this ratio across releases, we obtain insights into the dynamic relationship between code clone trends and bug-proneness, enhancing our understanding of code quality and maintenance over time.\\

\noindent\textbf{Investigating bug-proneness in "thin clones" vs. "thick clones."} We identify \textit{"thin clones"} and \textit{"thick clones"} based on the distribution of clones based on the percentiles. Firstly, we gather all the clone snapshots from all the frameworks. For each snapshot, the median cloned code size is computed, taking into account the diverse sizes of clone pairs within each snapshot of the codebase. Subsequently, the $25^{th}$ and $75^{th}$ percentiles are calculated, establishing thresholds for \textit{"thin clone"} and \textit{"thick clones"}, respectively. More specifically, code clones falling below the $25^{th}$ percentile are categorized as \textit{"thin clone"}, while those exceeding the $75^{th}$ percentile are labelled as \textit{"thick clone"}. Following the identification of bug-fixing commits, we map each change to the corresponding cloned code (as described in previous steps). Through this mapping, we categorize each commit-fix as either fixing a \textit{"thin clone"} or a \textit{"thick clone"}. To assess the statistical significance of bug-proneness differences between "thin clones" and "thick clones," we employ a Mann-Whitney U test on the percentage distribution of the number of changes to cloned code in \textit{"thin"} and \textit{"thick"} clones. This non-parametric test is chosen due to the potential non-normality of the data and its suitability for comparing two independent samples. If we obtain a p-value <= 0.05, we reject the null hypothesis and conclude that the distributions of bug-fixing percentages in the two code clone categories are different. Studying bug-proneness in "thin clones" versus "thick clones" provides insights into whether bugs tend to concentrate more on one type of clone over the other, which can help understand how code clones impact software quality and maintenance.\\

\noindent\textbf{Investigating the bug propagation in clones.} Different trends of code clones may exhibit diverse characteristics, such as cloned code size and distribution of thick and thin clones, which could influence how bugs propagate within the codebase. We formulated the following null hypothesis: \textit{H01: Different trends of code clones exhibit different bug propagation trends.} To investigate the bug propagation within code clones, we examine the number of file changes per bug-fixing commit, with the aim of discerning potential differences in bug propagation across various trends. Specifically, we evaluate whether the bug propagation varies significantly among different trends. 

To assess significance, we employ Welch's ANOVA test from scipy library \footnote{https://docs.scipy.org/doc/scipy/reference/generated/scipy.stats.f\_oneway.html}, a statistical analysis method suited for comparisons of groups with unequal sizes. Additionally, we also compare bug propagation, i.e., by examining the number of file changes per bug-fixing commit, in cloned and non-cloned code. This comprehensive approach allows us to gain insights into the distinctive patterns of bug propagation within code clones and explore potential variations in comparison to non-cloned code.\\

\noindent\textbf{Investigating the community size evolution in clones.} The community size is used to measure the proportion of contributors, i.e., authors of the code, engaged in clone-related activities within DL frameworks relative to the total contributors involved in the entire codebase. The community size represents an indicator of the collaborative involvement of contributors to clones. To derive the community size evolution, we employ the following steps. First, for each release, we calculate the number of contributors engaged in clone-related activities up to the release date. Subsequently, we determine the community size by dividing the number of contributors involved in clones by the total count of contributors to the entire codebase up to the respective release date. This metric provides a measure of how the number of contributors in clone-related efforts evolves over the course of the framework.

\subsubsection{Results}

\noindent\textbf{The decline in overall clone coverage observed in the \textit{"Decreasing"} and \textit{"Rise and Fall"} trends can be attributed to 1) a decrease in the cloned code size or a slow increase in the number of cloned code size compared to a rapid expansion of the codebase.} We conduct a comparison of cloned code size at the initial and final releases of each framework. As shown in Table \ref{tab:code_clone_size_analysis}, we observe that BentoML presents a reduction in cloned code size over time, resulting from a decrease in clone groups and clone siblings within a clone group. For example, despite a more than twofold increase in framework size between the first and last releases, BentoML's cloned code size decreased from 982 to 174 lines of code. This decrease in cloned code size corresponds to a decrease in the clone groups from 23 to 5. Hence, the decrease in clone coverage of BentoML from 10\% in the first release to less than 1\% in the last release. Aesara, Keras, and PyTorch also exhibit a decrease in clone coverage between the first and last release. Specifically, the clone coverage for Aesara, Keras, and PyTorch drop from 7\% to 5\%, 9\% to 4\% and 26\% to 2\%, respectively, between the first and last release. However, the decrease for the three frameworks is attributed to the rapid expansion of the codebase, outpacing the growth rate of the cloned code size. For example, while the codebase for Pytorch increased by a factor of 30 between the first and last release, the size of the clone code only increased by less than threefold.\\

\begin{table}[h]
\centering
\caption{Cloned code size analysis of the DL frameworks exhibiting "Decreasing" and "Rise and Fall" trends from first to last Release.}
\label{tab:code_clone_size_analysis}
\small
\begin{tabular}{l|l*{4}c}
\toprule
     & \textbf{Metric} &\textbf{First release}  & \textbf{Last release} & \textbf{Trend}\textbf{}\\ \midrule

\multirow{5}{*}{\makecell{BentoML}}
  & SLOC & 9,917 & 22,494 & $\nearrow$\\
    & Code size (SLOC) & 982 & 174 & $\searrow$\\
    & Clone coverage & 0.1 & 0.008 & $\searrow$\\
    & \# of siblings & 58 & 18 & $\searrow$\\
    & \# of groups & 23 & 11 & $\searrow$\\
 \bottomrule

\multirow{5}{*}{\makecell{Aesara}} & 
  SLOC & 39,021   & 84,773 & $\nearrow$\\
  & Code size (SLOC) & 2,765 & 4,612 & $\nearrow$\\
  & Clone coverage & 0.07 & 0.05 & $\searrow$\\
  & Total number of clone siblings & 307 & 396 & $\nearrow$\\
  & Total number of clone groups & 117 & 139 & $\nearrow$\\
 \bottomrule

\multirow{5}{*}{\makecell{Keras}} & 
  SLOC & 19,382   & 139,816 & $\nearrow$\\
  & Code size (SLOC) & 1,680 & 5,947 & $\nearrow$\\
  & Clone coverage & 0.09 & 0.04 & $\searrow$\\
  & \# of siblings & 117 & 335 & $\nearrow$\\
  & \# of groups & 43 & 98 & $\nearrow$\\
 \bottomrule

\multirow{5}{*}{\makecell{PyTorch}} & 
  SLOC & 10,567   & 313,603 & $\nearrow$\\
  & Code size (SLOC) & 2,741 & 7,527 & $\nearrow$\\
  & Clone coverage & 0.26 & 0.02 & $\searrow$\\
  & \# of siblings & 275 & 583 & $\nearrow$\\
  & \# of groups & 62 & 202 & $\nearrow$\\
 \bottomrule
 
\end{tabular}
\end{table}

\noindent\textbf{The decline in cloned code size can be attributed to code refactoring, third-party library reuse, and code clone removal associated with feature elimination.} To understand the factors contributing to the decrease in clone code density, we conduct an inter-release manual analysis by reading the release notes and commit messages for releases demonstrating a reduction in cloned code size compared to their preceding release. We find that cloned code size reduction is due to 1) code refactoring, 2) substitution of code fragments due to the use of third-party libraries, and 3) the removal of clone code as a consequence of eliminating certain features. For instance, duplicate code is moved to a shared utility code. For example, in release \textit{2.2.0}\footnote{https://github.com/keras-team/keras/releases/tag/2.2.0}, the Keras framework underwent a large code refactoring\footnote{https://github.com/keras-team/keras/pull/10865/commits} that affected the clones. In another release \textit{2.4.0}\footnote{https://github.com/keras-team/keras/releases/tag/2.4.0} of the Keras framework, the cloned code size is reduced due to a redirect of all APIs in the Keras package to tf.keras third party-library. In the release \textit{2.2.0}\footnote{https://github.com/aesara-devs/aesara/releases/tag/rel-2.2.0} of the Aesara framework, clones are eliminated as a result of the elimination of the function \textit{local\_neg\_neg} from the tensor module.\\

\begin{figure*}
\centering
\includegraphics[width=1\linewidth]{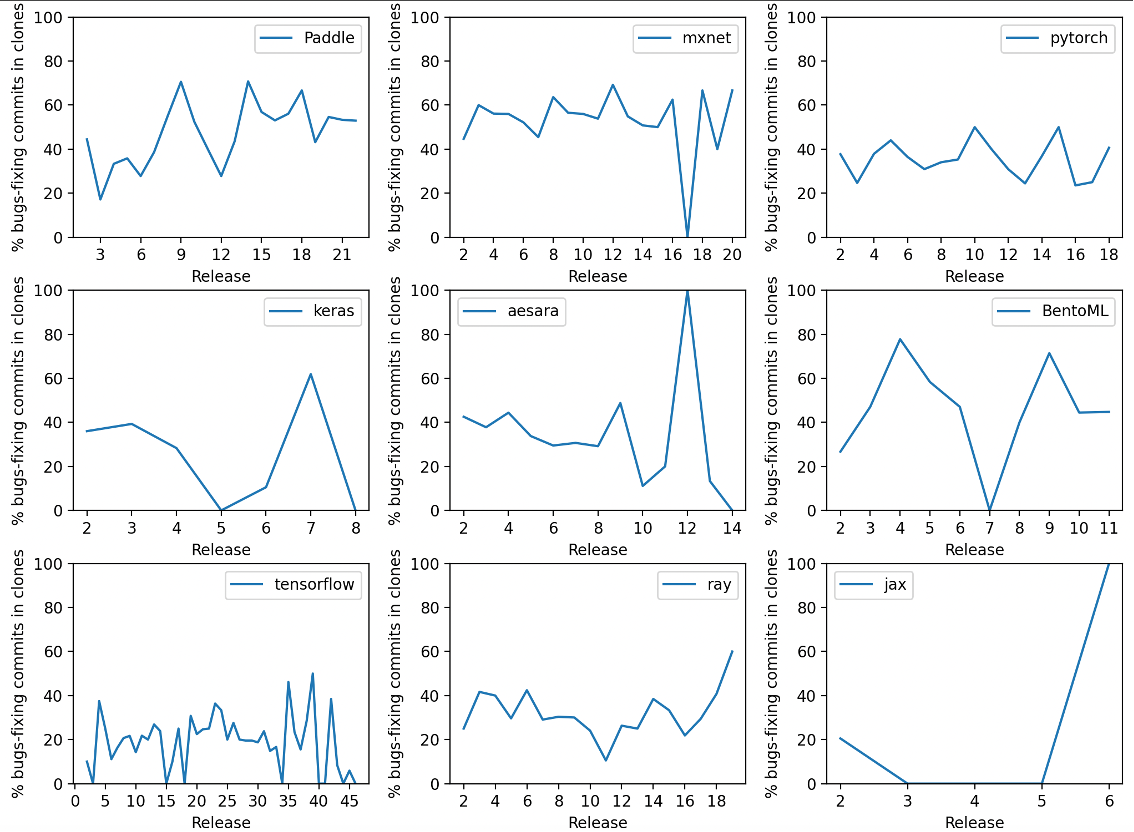}
\caption{The bug-proneness evolution in cloned code over releases.}
\label{fig:bug_evolution_trends}
\end{figure*}

\noindent\textbf{Bug fixing is a persistent activity consistently occurring throughout the lifespan of frameworks, among all the code cloning trends. The \textit{"Serpentine"} trend is more susceptible to bugs.} Figure \ref{fig:bug_evolution_trends} represents the evolution of bug-proneness in clone code in DL frameworks over releases. When comparing bug-proneness percentages across frameworks, we notice distinct variations. For instance, frameworks belonging to the \textit{"Serpentine"} trend, i.e., \textit{Paddle} and \textit{MXNet}, demonstrate higher bug-proneness percentages across releases in comparison to the frameworks belonging to the \textit{"Decreasing"}, \textit{"Rise and Fall"} and \textit{"Stable} trends. For example, \textit{Paddle} and \textit{MXNet} frameworks have 50\% and 75\% of the releases having more than 50\% bug-fixing commits in clones respectively, whereas in \textit{TensorFlow} all 100\% of the framework releases have less than 50\% of the bug-fixing commits in clones as shown in  \ref{fig:bug_evolution_trends}. ANOVA-Welch test indicates a significant difference between the bug-proneness evolution of \textit{Paddle} and \textit{MXNet} frameworks and the frameworks belonging to the other three trends. These findings suggest that the DL frameworks belonging to the \textit{"Serpentine"} trend characterized by a fluctuating code clone trend could be more susceptible to bugs. 

Moreover, the \textit{Keras} and \textit{Aesara} frameworks belonging to the \textit{"Rise and Fall"} clone trend exhibit a decreasing bug-proneness trend where the overall bug-proneness percentage decreases from approximately 40\% in initial releases to almost zero in subsequent releases. For the Aesara framework, we notice a spike to 100\% in the bug-proneness at release 12, meaning that all of the commits introduced were fixing bugs. Further investigation demonstrates that the 100\% rate is due to a short release cycle of 11 days, during which only one commit to a clone occurs, and this solitary commit happens to be a bug fix. Similarly, the abrupt decrease in bug-proneness to zero in the \textit{MXNet} framework at release 17 is linked to the absence of commits to code clones during a relatively short release cycle of 27 days. The observed descending trends in code clones, coupled with a corresponding reduction in bug-proneness, suggest an enhancement in overall code quality and reliability. These findings underscore the importance of investigating the coding practices followed within the release and their impacts on the code clone in DL frameworks, which we address in the following research question.\\

\begin{figure*}
\centering
\includegraphics[width=1\linewidth]{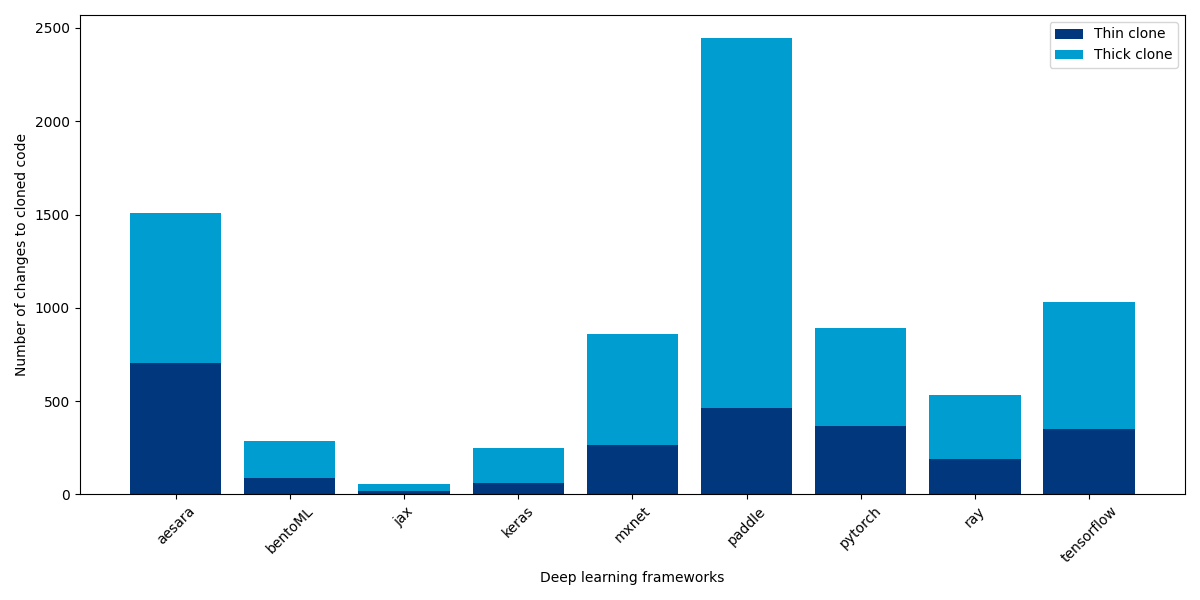}
\caption{Bug-fixing commits distribution by \textit{"thin clones"} and \textit{"thick clones"}.}
\label{fig:thin_thick_results}
\end{figure*}

\noindent\textbf{Over 50\% of the changes implemented in bug-fixing commits predominantly occur within \textit{"thick"} cloned code as opposed to \textit{"thin"} cloned code across all the long-term code clone trends.} Figure \ref{fig:thin_thick_results} shows the distribution of changes across the frameworks. The Mann-Whitney U test reveals a significant difference in bug-proneness between \textit{"thin} clones and \textit{"thick"} clones with p-value < 0.05. In addition, we investigate if the difference among the clone categories is of strong significance by calculating the Common Rank (Z) score. We obtain a Z value of 3.57 that suggests a substantial and statistically significant difference between the two bug-fixing commits in \textit{"thick"} and \textit{"thin"} code clones. A z-score \footnote{https://www.z-table.com/} of +/- 1.96 or greater is considered statistically significant at the 5\% level of significance (i.e., p < 0.05). Additionally, we employ the Cliff's Delta as another measure of effect size. A value of 1 indicates a large effect size. We obtain a  Cliff's Delta value of 1, which further reinforces the notion that the bug-proneness difference between \textit{"thin"} and \textit{"thick"}. These results provide statistical evidence supporting the assertion that bug-proneness varies significantly between \textit{"thin} clones and \textit{"thick"} clones in our study. This result highlights a substantial distinction in bug-fixing tendencies between the two clone categories with \textit{"thick"} clones being more susceptible to bug-fixing commits.\\

\noindent\textbf{Our analysis reveals no significant difference in the number of files changed per bug-fixing commit across long-term code clone trends, as validated by Welch's ANOVA test, and we reject the null hypothesis.} However, comparing the bug propagation between cloned and non-cloned code across the frameworks indicates a significant difference in the number of files changed per bug-fixing commit in clones versus non-clones. Specifically, non-cloned files exhibit a higher number of changes, with an average of 2.4 files changed per commit, compared to 1.5 files changed per commit in cloned code. This could suggest that bug fixes in cloned sections are more localized and targeted. \\

\begin{figure*}
\centering
\includegraphics[width=1\linewidth]{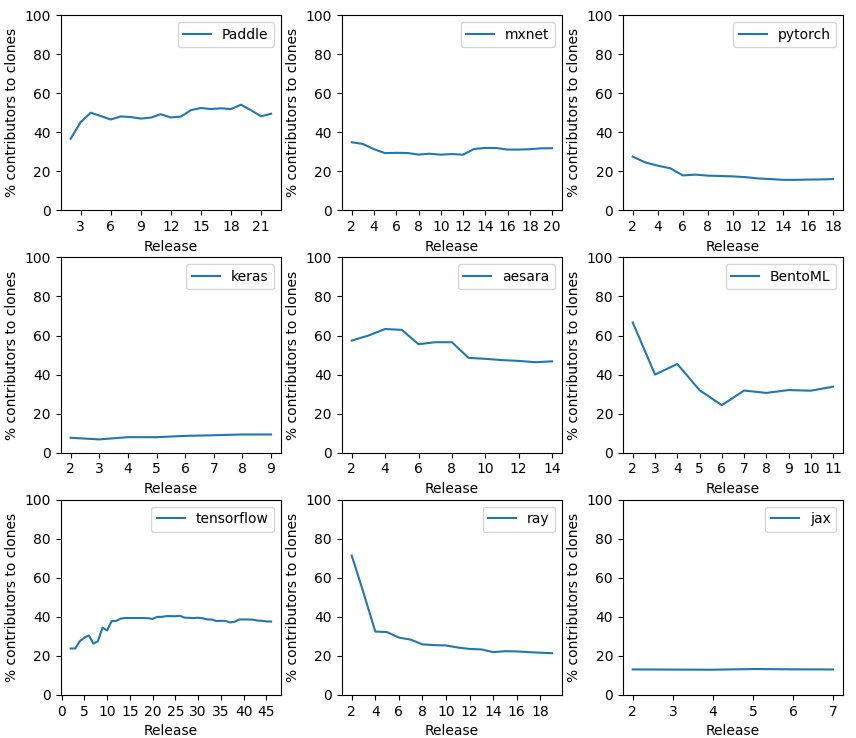}
\caption{The evolution of code clone community size over releases.}
\label{fig:contributors_to_clones}
\end{figure*}

\noindent\textbf{The community involvement in clone activities remains consistent throughout the lifetime of the frameworks, with exceptions in \textit{BentoML} and \textit{Ray}, where original authors dominate clone maintenance and the community size witnesses a decrease over time.} For instance, a relatively small portion of the community actively contributes to clones, consistently amounting to less than 50\% for the majority of releases. Figure \ref{fig:contributors_to_clones} depicts the community size evolution, i.e., the percentage of contributors to clones across various DL frameworks. As we can notice, the percentage of the community involved in clone-related presents a stable and sustained level of contribution to clones over time, among all the long-term cloning trends of the framework. However, there is a distinct reduction in community involvement in code clones within the frameworks \textit{BentoML} and \textit{Ray}. This decline in community participation can be attributed to the circumstance where the contributors making changes to clones are predominantly the original authors of those clones. To investigate this observation further, we manually check the clone pairs within \textit{BentoML} and \textit{Ray}. We notice that for those frameworks, the original creators of the clones are the ones maintaining them. 

\begin{Summary}{}{firstsummary} Long-term code cloning trends exhibit some common and distinct characteristics. The decline in overall clone coverage observed in the \textit{"Decreasing"} and \textit{"Rise and Fall"} trends is attributed to 1) a decrease in the cloned code size or 2) a slow increase in the cloned code size compared to a rapid expansion of the codebase. All long-term trends are prone to bugs, and the bug-fixing activities are consistent across the lifespan of the frameworks of all the trends. However, the framework belonging to the \textit{"Serpentine"} trend, i.e., \textit{Paddle} and \textit{MXNet}, could be more susceptible to bugs as they present a higher percentage of bug-proneness across the releases indicated by a higher commit-fixing. \textit{"Thick"} clones present a higher bug-proneness as compared to \textit{"thin"} clones across all the long-term trends. Regarding the community involvement in clone activities, it remains consistent through the lifetime of the frameworks, with exceptions in \textit{BentoML} and \textit{Ray}, where original authors dominate clone maintenance and the community size witnesses a decrease over time.
\end{Summary}

\subsection{RQ2: What are the characteristics of within-release development patterns and do these patterns contribute to the overarching long-term trends in code cloning?}

\label{sec:RQ2}
\subsubsection{Motivation} In RQ1, we have investigated the characteristics of long-term clone trends over releases and observed that every trend has distinct characteristics, such as high susceptibility of the \textit{"Serpentine"} trend to bug-proneness. In this RQ, we are interested in investigating the short-term clone patterns within each individual release. Our goal is to discern the impact of these patterns on long-term cloning trends and identify the characteristics of the within-release code cloning patterns. More specifically, we want to understand the factors influencing the evolution of cloned code size and provide insights into the dynamics of development practices within individual releases. In addition, the gained knowledge about within-release patterns, when aggregated over multiple releases, could contribute to understanding how development practices within releases influence clone evolution trends over time. By uncovering these relationships, we seek to formulate insights to enhance the efficiency and maintainability of DL frameworks within the code clone context.

\subsubsection{Approach} 

After we obtain the three within-release code cloning patterns \textit{"Ascending"}, \textit{"Descending"}, and \textit{"Steady"}, we aim to obtain a chronological evolution of how the within-release pattern contributed to the long-term trends. Therefore, we construct, for every DL framework, the evolution sequence of within-release patterns over the framework releases. Table \ref{tab:release_development_trends} depicts the obtained chronological pattern sequences. \textit{$P_a$}, \textit{$P_d$} and \textit{$P_s$} denotes an \textit{"Ascending"}, \textit{"Descending"}, and \textit{"Steady"} pattern respectively.\\

Our approach to building the regression models comprises the following three steps:

\begin{itemize}[label=--]
  \item \textbf{Step1: Correlation and redundancy analysis of the independent variables.} we collect 37 metrics as detailed in section \ref{sec:metric_calculation}. The presence of correlated metrics might affect the performance of the model \cite{10.1007/s10664-020-09848-1}, therefore, we apply varclus\footnote{https://search.r-project.org/CRAN/refmans/Hmisc/html/varclus.html} function in R to detect the existence of collinearity and exclude one metric of any pair of metrics that achieves a coefficient of 0.7 \cite{10.1007/s10664-015-9381-9}. Then, we calculate the Variance Inflation Factors (VIFs) using the VIF\footnote{https://www.rdocumentation.org/packages/regclass/versions/1.6/topics/VIF} function in R for each independent variable to detect multicollinearity issues. We exclude all independent variables with VIF values above 5 because it indicates the existence of multicollinearity \cite{cohen2002applied, weisberg2005applied}. 
\\

\item \textbf{Step 2: Constructing the models.} We construct three different regression models for each pattern shown in Figure \ref{fig:within_release_patterns}: $Model_{Ascending}$, $Model_{Descending}$, and $Model_{Steady}$. For a $Model_i$, we assign to each snapshot revision, i.e., commit, the value 1 if the associated within-release pattern is \textit{i} and 0 otherwise. For example, for $Model_{Ascending}$, we assign 1 to the commits that are clustered in the release belonging to the \textit{"Ascending"} pattern. Since our dependent variable is binary, we follow the existing work \cite{Yang2022AnES, Khoshgoftaar1999LOGISTICRM, 10.1145/1368088.1368160} and employ logistic regression models. In particular, we fit a binomial logistic regression model using the glm\footnote{https://www.rdocumentation.org/packages/stats/versions/3.6.2/topics/glm} function provided by the state R package.\\
  
\item \textbf{Step 3: Evaluating the models and assessing the significance of the independent variables.} After constructing the models, we use the Area Under the Receiver Operating Characteristic Curve (AUC) \cite{Ling2003AUCAB} to assess the predictive power of the created logistic models. A predictive model is deemed promising if the AUC-ROC is 0.7 or higher, with 1 signifying perfect predictive power\cite{gorunescu2011data, hanley1982meaning}. Next, we employ the ANOVA test \cite{pinheiro_link} to assess the significance, measured in terms of (\(\chi^2\)) for each independent variable in our model.
  The (\(\chi^2\)) values for each variable are calculated as a percentage of the total (\(\chi^2\)) values for all variables. We use upward and downward arrows to signify direct and inverse relationships between independent and dependent variables, respectively.

\end{itemize}

\subsection{Results}

\textbf{Long-term descending trends, i.e., \textit{"Decreasing"} and \textit{"Rise and Fall"}, in addition to \textit{"Stable"} trends consistently exhibit \textit{"Steady"} within-release patterns}. As we notice in table \ref{tab:release_development_trends}, the frameworks belonging to the long-term descending trends, i.e., \textit{Keras}, \textit{Aesara}, \textit{BentoML}, \textit{PyTorch}, \textit{JAX}, \textit{Ray} and \textit{TensorFlow} are characterized by a consistent and consecutive repetitive stable pattern \textit{P$_{S}$}. These stable patterns constitute over 30\% of the total within-release patterns of each framework. This observation shows a connection between the two temporal scales and how the patterns of code clones within releases, i.e., short-term, have long-term implications. Hence, we proceed next with understanding the characteristics of the within-release code clone development practices.\\

\begin{table}[h]
\centering
\caption{Within-release patterns. \textit{$P_a$}, \textit{$P_d$} and \textit{$P_s$} denote an \textit{"Ascending"}, \textit{"Descending"}, and \textit{"Steady"} pattern respectively.}
\label{tab:release_development_trends}
\small
\begin{tabular}{ll}

\toprule
     \textbf{DL framework} &\textbf{Within-release chronological patterns}\\ \midrule
  MXNet & \(
\textcolor{red}{P_{a}}, \textcolor{red}{P_{a}}, \textcolor{red}{P_{a}}, \textcolor{green}{P_{d}}, \textcolor{green}{P_{d}}, \textcolor{green}{P_{d}}, \textcolor{green}{P_{d}}, \textcolor{green}{P_{d}}, \textcolor{green}{P_{d}}, \textcolor{green}{P_{d}}, \textcolor{red}{P_{a}}, \textcolor{red}{P_{a}}, \textcolor{red}{P_{a}}, \textcolor{red}{P_{a}}, \textcolor{red}{P_{a}}, \textcolor{red}{P_{a}}, \textcolor{red}{P_{a}}, \textcolor{red}{P_{a}}, \textcolor{green}{P_{d}}, \textcolor{green}{P_{d}}
\)\\
     Paddle & \(
\textcolor{blue}{P_{s}}, \textcolor{red}{P_{a}}, \textcolor{red}{P_{a}}, \textcolor{red}{P_{a}}, \textcolor{green}{P_{d}}, \textcolor{green}{P_{d}}, \textcolor{green}{P_{d}}, \textcolor{green}{P_{d}}, \textcolor{green}{P_{d}}, \textcolor{green}{P_{d}}, \textcolor{green}{P_{d}}, \textcolor{red}{P_{a}}, \textcolor{green}{P_{d}}, \textcolor{red}{P_{a}}, \textcolor{blue}{P_{s}}, \textcolor{red}{P_{a}}, \textcolor{red}{P_{a}}, \textcolor{red}{P_{a}}, \textcolor{red}{P_{a}}, \textcolor{red}{P_{a}}, \textcolor{red}{P_{a}}
\) \\
 Keras & \(
\textcolor{red}{P_{a}}, \textcolor{red}{P_{a}}, \textcolor{red}{P_{a}}, \textcolor{red}{P_{a}}, \textcolor{blue}{P_{s}}, \textcolor{blue}{P_{s}}, \textcolor{blue}{P_{s}}, \textcolor{blue}{P_{s}}
\)\\
  Aesara & \(
\textcolor{red}{P_{a}}, \textcolor{red}{P_{a}}, \textcolor{red}{P_{a}}, \textcolor{red}{P_{a}}, \textcolor{red}{P_{a}}, \textcolor{red}{P_{a}}, \textcolor{red}{P_{a}}, \textcolor{red}{P_{a}}, \textcolor{blue}{P_{s}}, \textcolor{blue}{P_{s}}, \textcolor{blue}{P_{s}}, \textcolor{blue}{P_{s}}
\)\\
  BentoML & \(
\textcolor{red}{P_{a}}, \textcolor{red}{P_{a}}, \textcolor{green}{P_{d}}, \textcolor{green}{P_{d}}, \textcolor{green}{P_{d}}, \textcolor{green}{P_{d}}, \textcolor{red}{P_{a}}, \textcolor{red}{P_{a}}, \textcolor{blue}{P_{s}}, \textcolor{blue}{P_{s}}, \textcolor{blue}{P_{s}}
\)\\
  PyTorch & \(
\textcolor{green}{P_{d}}, \textcolor{green}{P_{d}}, \textcolor{green}{P_{d}}, \textcolor{green}{P_{d}}, \textcolor{blue}{P_{s}}, \textcolor{blue}{P_{s}}, \textcolor{blue}{P_{s}}, \textcolor{blue}{P_{s}}, \textcolor{blue}{P_{s}}, \textcolor{blue}{P_{s}}, \textcolor{blue}{P_{s}}, \textcolor{blue}{P_{s}}, \textcolor{blue}{P_{s}}, \textcolor{blue}{P_{s}}, \textcolor{blue}{P_{s}}, \textcolor{blue}{P_{s}}, \textcolor{blue}{P_{s}}, \textcolor{blue}{P_{s}}
\)\\  
 JAX & \(
\textcolor{blue}{P_{s}}, \textcolor{blue}{P_{s}}, \textcolor{blue}{P_{s}}, \textcolor{blue}{P_{s}}, \textcolor{blue}{P_{s}}
\)\\
 Ray & \(
\textcolor{blue}{P_{s}}, \textcolor{blue}{P_{s}}, \textcolor{blue}{P_{s}}, \textcolor{blue}{P_{s}}, \textcolor{blue}{P_{s}}, \textcolor{blue}{P_{s}}, \textcolor{blue}{P_{s}}, \textcolor{blue}{P_{s}}, \textcolor{blue}{P_{s}}, \textcolor{blue}{P_{s}}, \textcolor{blue}{P_{s}}, \textcolor{blue}{P_{s}}, \textcolor{blue}{P_{s}}, \textcolor{blue}{P_{s}}, \textcolor{blue}{P_{s}}, \textcolor{blue}{P_{s}}, \textcolor{blue}{P_{s}}, \textcolor{blue}{P_{s}}, \textcolor{blue}{P_{s}}
\)\\
 TensforFlow & \(
\textcolor{blue}{P_{s}}, \textcolor{blue}{P_{s}}, \textcolor{blue}{P_{s}}, \textcolor{blue}{P_{s}}, \textcolor{blue}{P_{s}}, \textcolor{blue}{P_{s}}, \textcolor{blue}{P_{s}}, \textcolor{blue}{P_{s}}, \textcolor{blue}{P_{s}}, \textcolor{blue}{P_{s}}, \textcolor{blue}{P_{s}}, \textcolor{blue}{P_{s}}, \textcolor{blue}{P_{s}}, \textcolor{blue}{P_{s}}, \textcolor{blue}{P_{s}}, \textcolor{blue}{P_{s}}, \textcolor{blue}{P_{s}}, \textcolor{blue}{P_{s}}, \textcolor{blue}{P_{s}}, \textcolor{blue}{P_{s}}, \textcolor{blue}{P_{s}},
\textcolor{blue}{P_{s}}, 
\)\\
& \(
\textcolor{blue}{P_{s}}, \textcolor{blue}{P_{s}}, \textcolor{blue}{P_{s}}, \textcolor{blue}{P_{s}}, \textcolor{blue}{P_{s}}, \textcolor{blue}{P_{s}}, \textcolor{blue}{P_{s}}, \textcolor{blue}{P_{s}}, \textcolor{blue}{P_{s}}, \textcolor{blue}{P_{s}}, \textcolor{blue}{P_{s}}, \textcolor{blue}{P_{s}}, \textcolor{blue}{P_{s}}, \textcolor{blue}{P_{s}}, \textcolor{blue}{P_{s}}, \textcolor{blue}{P_{s}}\)\\
 \bottomrule
\end{tabular}
\end{table}

\noindent\textbf{The $\bm{Model_{Descending}}$ and $\bm{Model_{Steady}}$ demonstrate the most robust explanatory capability among the three constructed models.} Table 10 provides an overview of our created models. The logistic regression models for $Model_{Descending}$ and $Model_{Steady}$ exhibit the highest AUC values at 0.92, whereas $Model_{Ascending}$ achieved an AUC of 0.89. Each model highlights a distinct attribute with the greatest explanatory strength, as shown in Table \ref{tab:lr_model_results} (due to the space constraint, the table shows only the top five significant features for all three constructed models). Further insights into the analysis of these constructed models are detailed below.\\

\begin{table}[h]
    \centering
    \caption{A summary of the analysis of the constructed three regression models.}
    \label{tab:lr_model_results}
    \small
    \begin{tabular}{lccccc}
        \toprule
        \textbf{Metrics} & \textbf{Coef.} & \textbf{\(\bm{\chi^2}\)(\%)} & \textbf{Pr(<\(\bm{\chi^2}\))} & 
        \textbf{Sign.} & \textbf{Relationship} \\ \midrule
         & \multicolumn{5}{c}{\textit{Model$_{\text{Ascending}}$}} \\  \midrule
         \# of committers to a clone pair & -3.62 & 39.67 & $< 2.2e^{-16}$  & *** & $\searrow$ \\
         complexity of abstract module & 7.74 & 23.90 & $< 2.2e^{-16}$  & *** & $\nearrow$ \\
         size of cloned code (med) & 4.98 & 21.56 & $< 2.2e^{-16}$  & *** & $\nearrow$ \\
         \# clones siblings (max) & -5.30 & 16.99 & $< 2.2e^{-16}$  & *** & $\searrow$ \\
         clone life (\# commits) & 9.50 & 4.85 & $< 2.2e^{-16}$  & *** & $\nearrow$ \\
        \midrule
         & \multicolumn{5}{c}{\textit{Model$_{\text{Descending}}$}} 
         \\  \midrule  
          \# changes to clone pairs & -9.08 & 24.86 & $< 2.2e^{-16}$  & *** & $\searrow$ \\
          complexity of abstract module & -18.58 & 24.63 & $< 2.2e^{-16}$  & *** & $\searrow$ \\
          size of cloned code (med) & 8.00 & 23.83 & $< 2.2e^{-16}$  & *** & $\nearrow$ \\
          similarity & -5.18 & 14.58 & $< 2.2e^{-16}$ & *** & $\searrow$ \\
          cyclomatic complexity & 3.78 & 11.94 & $< 2.2e^{-16}$ & *** & $\nearrow$ \\
          \midrule
          & \multicolumn{5}{c}{\textit{Model$_{\text{Steady}}$}} \\  \midrule  
          \# of declaration statement (med) & -8.54 & 36.56 & $< 2.2e^{-16}$  & *** & $\searrow$ \\
         \# abstract classes (med) & 4.36 & 25.61 & $< 2.2e^{-16}$  & *** & $\nearrow$ \\
         \# changes to clone pairs & 3.34 & 18.08 & $< 2.2e^{-16}$  & *** & $\nearrow$ \\
         \# of clones siblings (med) &  45.89 & 6.85 & $< 2.2e^{-16}$  & *** & $\nearrow$ \\
         \# of unique file & -2.70 & 5.78 & $< 2.2e^{-16}$ & *** & $\searrow$ \\
        \bottomrule
    \end{tabular}
\end{table}

\noindent\textbf{"Ascending" code cloning pattern analysis}

\begin{itemize}[label=--]
    \item \noindent\textbf{The decreased involvement of committers in the clone pairs clone genealogy is associated with a larger cloned code size}. As shown in Table \ref{tab:lr_model_results}, the feature number of committers to a clone pairs accounts for the highest (\(\chi^2\)) in the $Model_{Ascending}$, suggesting that the fewer is the number of developers changing a clone pair, the higher is the likelihood of cloned code size increase. This could be explained by the fact that fewer committers modifying the clone pairs know the code well and are propagating the changes while maintaining the code clones. When new committers make changes to existing clones, the probability of applying the changes to all clones' copies decreases as they might not be aware of its presence, 
    leading to inconsistent copies, hence a decreased cloned code size. For example, in the \textit{PyTorch} framework, the cloned functions \textit{acc\_ops\_max\_pool2d} and \textit{acc\_ops\_avg\_pool2d} of the converter module were modified consistently and maintained the consistent state through the three commits \textit{239b38268b\footnote{https://github.com/pytorch/pytorch/commit/239b38268b}},\textit{0d8a8a2e41\footnote{https://github.com/pytorch/pytorch/commit/0d8a8a2e41}} and  \textit{e23827e6d6\footnote{https://github.com/pytorch/pytorch/commit/e23827e6d6}} by the same committer. Whereas, in another instance of the same framework, the optimizer \textit{AdagradOptimizer} and \textit{AdamOptimizer} that were clones were modified independently. A first commit \textit{05542f6222\footnote{https://github.com/pytorch/pytorch/commit/05542f6222}} made by one committer made changes to \textit{AdagradOptimizer} only leading to an inconsistent state, then another commit by another committer \textit{812bc1dde6\footnote{https://github.com/pytorch/pytorch/commit/812bc1dde6}} propagated the change to the second optimizer, which led a consistent state again. \textbf{In addition, we highlight that the positive correlation (in the same model) between clone pairs' longevity (i.e., clone life in terms of the number of commits and the increased cloned code size) suggests that the clone pairs are either changed consistently or not modified}.\\
    
    \item \noindent\textbf{Code clones size tends to increase when abstract classes exhibit higher complexity}. Cloned code size positively correlates to the complexity of the cloned code in abstract classes. It is the second most explanatory factor in $Model_{Ascending}$. The rationale is likely rooted in the observation that code cloning often occurs under conditions of higher code complexity within a class. When a class is more complex, it may contain a greater number of methods, variables, or functionalities. In such intricate classes, developers might encounter challenges related to code reuse or modularization. As a result, they may resort to code cloning as a quick solution. This also aligned with the fact that the increased cloned code size is positively correlated with a larger cloned code size as it is the third most significant explanatory factor, as shown in Table \ref{tab:lr_model_results}.\\
\end{itemize}

\noindent\textbf{"Descending" code cloning pattern analysis}

\begin{itemize}[label=--]
    \item \noindent\textbf{Code clones pairs in \textit{"Descending"} cloned code size pattern exhibit fewer changes}. The $Model_{Descending}$ indicates a significant association between the number of changes to clone pairs and the cloned code size. When fewer modifications are made to existing pairs of code clones, it suggests that these segments of code remain relatively stable over time. This stability implies that developers are refrained from making substantial alterations or additions to the existing code clones. \\
    
    \item \noindent\textbf{Reduced complexity in abstract classes is associated with less reliance on code cloning}. As depicted in Table \ref{tab:lr_model_results}, a reverse association exists between the complexity of the abstract module where the code clone is originated and the cloned code size. This finding aligns with the results obtained in $Model_{Ascending}$ where an increased complexity is associated with the increased cloned code size. This suggests developers probably need to engage in less copying and pasting of code when dealing with simplified code structures in abstract classes.\\
\end{itemize}

\noindent\textbf{\textbf{"Steady"} code cloning pattern analysis}
\begin{itemize}[label=--]
    \item \noindent\textbf{Code optimization strategies, such as reducing the number of declaration statements and incorporating modular structures, are significantly linked to cloned code size in "Steady" cloned code patterns}. Table 5 shows an inverse association between the number of declaration statements and the stable cloned code size in $Model_{Steady}$. A lower count of declaration statements within code clones suggests a more concise and modular structure, indicating a potential effective maintenance strategy where similar functionalities across the code base share the same set of globally declared variables. This finding aligns with the relevance observed in the descending code clone pattern analysis, where an increased number of abstract classes, i.e., the second most significant factor, is also associated with stable cloned code sizes. Smaller and modular code segments are commonly linked to better maintainability and ease of understanding, highlighting the significance of cohesive code structuring practices.\\
\end{itemize}

\begin{Summary}{}{firstsummary}
 Within-release code cloning patterns impact the long-term code cloning trends. For instance, long-term descending trends, i.e., "Decreasing" and "Rise and Fall," in addition to "Stable" trends, consistently exhibit "Steady" within-release patterns. Within-release patterns also exhibit distinct characteristics. Our results demonstrate that \textit{"Ascending"} code cloning pattern is associated with decreased committer involvement in clone pairs and increased cloned code size, suggesting that fewer committers may lead to a higher likelihood of cloned code size increase. In the \textit{"Descending"} pattern, clone pairs with reduced cloned code size exhibit fewer changes, indicating stability over time, and simplified code structures in abstract classes are associated with less reliance on code cloning. Lastly, the \textit{"Steady"} pattern links code optimization and smaller declaration statements count to stable cloned code sizes, emphasizing the significance of cohesive code structuring practices. Our work shows the characteristics of the within-release clone evolution, which helps developers prioritize their actions based on their within-release pattern.
\end{Summary}

\subsection{RQ3: How do code clones manifest and evolve across different DL frameworks?}
\label{sec:RQ3}
\subsubsection{Motivation} Different DL frameworks support distinct features and adopt different design philosophies. For example, while \textit{Jax} focuses on high-performance numerical computing, \textit{PyTorch} is known for its dynamic computation graph. However, some DL frameworks share common functionalities. For example, \textit{TensorFlow}, \textit{PyTorch}, and \textit{MXNet} exhibit overlapping functionalities, such as in neural network definition and training. The convergence of functionalities among these frameworks can result in comparable code patterns for similar tasks, potentially leading to code clones, as developers may employ similar constructs and operations. In this RQ, we conduct a cross-framework clone detection to identify and analyze similarities in code across different DL frameworks, and we track the evolution of the cloned functionalities. This is crucial for understanding how common functionalities of code clones manifest in DL, aiding in the development of standardized practices and fostering collaborative initiatives within the DL community.

\subsubsection{Approach}
The investigation process of cross-framework code clones involves the following steps:\\

\noindent\textbf{Step 1: Building quarterly snapshots and detecting cross-framework code clones.} In this step, we organize the commits of each DL framework into quarterly groups and select snapshots corresponding to the latest commit in each group for analysis. We choose the quarterly interval as it provides a fine-grained analysis of the cross-framework clone evolution. Then, we run the clone detection on the snapshots of the frameworks corresponding to the selected commits. As explained in section \ref{sec:clone_detection}, we use SourcererCC to detect the cross-framework clones every quarter, generating a comprehensive list of cross-framework clone file pairs for every quarter.\\

\noindent\textbf{Step 2: Constructing the time series for the evolution of cross-framework clones.} Following the cross-framework clone detection, we construct a time series for the quarterly evolution of cross-framework clones. This time series consists of data points representing the count of cross-framework clone files on specific dates. It spans 28 quarters, offering a dynamic record of how cross-framework clones evolve across DL frameworks over time.\\

\noindent\textbf{Step 3: Investigating cross-framework file-level clone pairs manually.} To provide deeper insights into the evolution of cross-framework clones in DL frameworks, we conduct a manual analysis. In particular, we manually investigate the identified cross-framework file-level clone pairs for each snapshot. First, we examine for each file clone the \textit{repository location}, such as the path or module, to identify the context in which the clones exist. Secondly, for every file pair, we check the \textit{source code} to identify the functionality of these files. Lastly, we track the \textit{code-changing activities} related to these clone files to better understand how they evolve over time and the reason they cease to be clones.

\subsubsection{Results} 

\begin{figure*}
\centering
\includegraphics[width=1\linewidth]{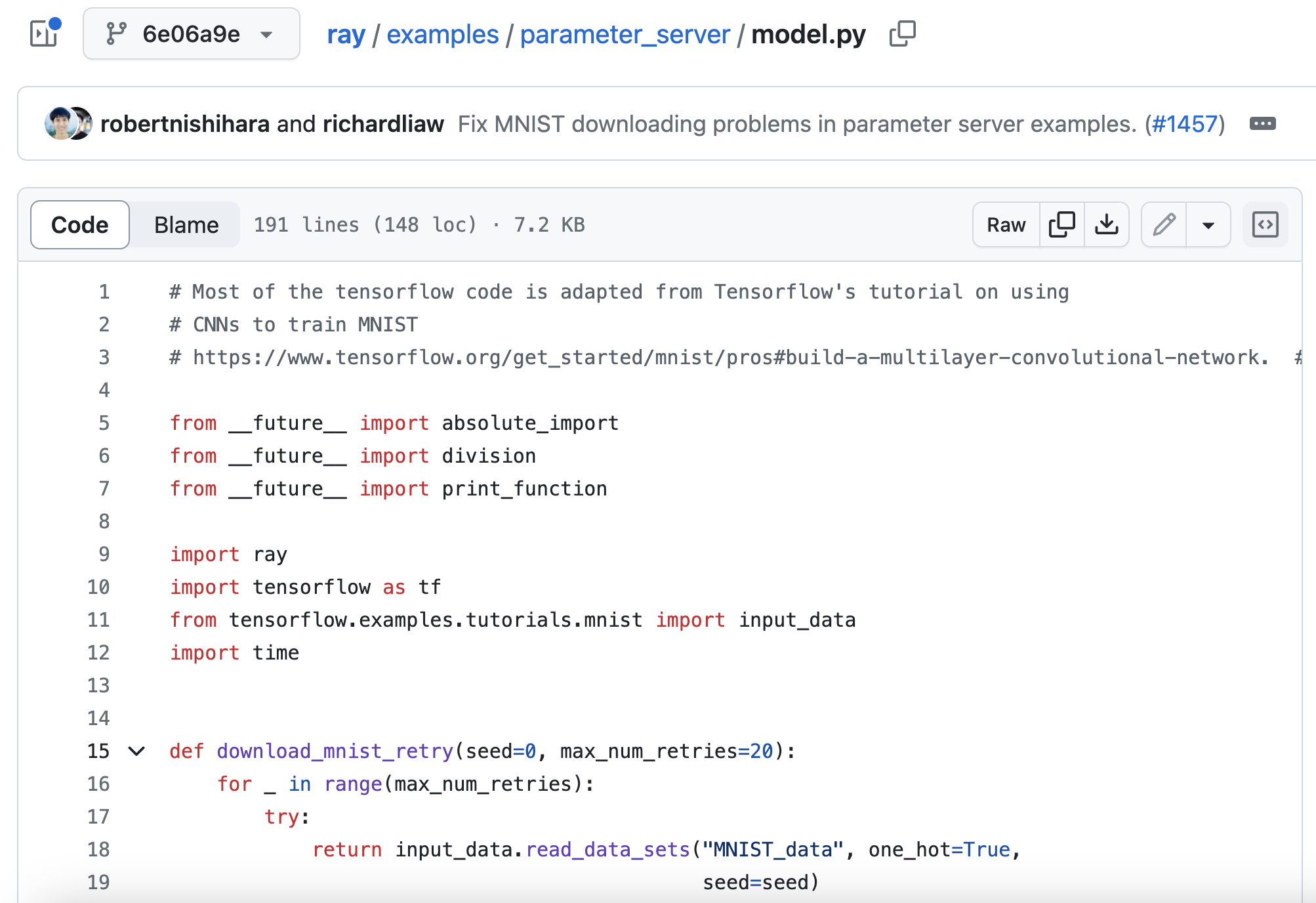}
\caption{Example of a functional code clone instance from TensorFlow to Ray.}
\label{functional_code_clone}
\end{figure*}

\noindent\textbf{DL frameworks present two main categories of cross-framework file-level code clones: the \textit{functional code clones} and the \textit{architectural adaptation clones}.} \textit{Functional code clones} include the clones where specific files, pertaining to distinct functionalities, are replicated or adapted within a DL framework. Figure \ref{functional_code_clone} illustrates an example \footnote{https://github.com/ray-project/ray/blob/6e06a9e338e1045fa0ba73b366bb78a2c7f0fef8/examples/parameter\_server/model.py} of the clone belonging to this category where some code was adapted from a TensorFlow tutorial for training CNNs on the MNIST dataset to be included in the \textit{Ray} framework. \textit{Architectural adaptation clones} represent the adaptation and integration of an entire module from one DL framework into another resulting in an architectural adaptation between frameworks. For example, we notice that the increase in clones in March 2018 was due to the integration of the \textit{Keras} module into \textit{TensorFlow}. In particular, the layer and applications modules of Keras were integrated into TensorFlow as exact duplication. We also observe that even though both categories of clones are present in DL frameworks, the \textit{architectural adaptation clones} represent the larger category among all file cross clones in DL frameworks.\\

\noindent\textbf{Functional code clones in DL frameworks belong to seven main categories \textit{Communication Interface, Distributed Training in TensorFlow, Python Object Serialization, Efficiency in Python, Version Control, Deep Learning Architectures,} and \textit{Probability and Statistics}}. We describe below each of the categories, and we include in Table \ref{tab:functional_code_clone_distribution} the distribution of the file clone pairs across the seven categories with the functionalities of the existing clones and the range of the lifetime of the file clone pairs.

\begin{itemize}[label=--]
  \item \textbf{Communication Interface:} Clones related to how DL frameworks handle data exchange or interaction, such as the protocol buffers (protobuf) messages and services for communication with a \textit{BentoML} repository.
  \item \textbf{Distributed Training in TensorFlow:} Clones associated with distributed training in the \textit{TensorFlow} DL framework, such as the gradient reduction (all-reduce) in distributed training using TensorFlow.
  \item \textbf{Python Object Serialization:} Clones related to the serialization of Python objects such as CloudPickle serialization.
  \item \textbf{Efficiency in Python:} Clones related to enhancing efficiency in Python, such as the LazyLoader class.
  \item \textbf{Version Control:} Clones pertaining to version control mechanisms, to manage and track changes in code versions such as Versioneer.
  \item \textbf{Deep Learning Architectures:} Clones related to deep learning architectures, suggesting shared design patterns and functionalities, such as Convolutional Neural Networks (CNN).
  \item \textbf{Probability and Statistics:} Clones related to probability and statistics, revealing common approaches in implementing statistical concepts, such as normal distribution.
\end{itemize}

As we notice, \textit{Probability and Statistics} and \textit{Deep Learning Architectures} represent the top two common categories constituting approximately 70\% of the file clone pairs. This insight suggests an overlap in specific functionalities and design patterns across different DL frameworks, emphasizing commonalities in the implementation of key features such as Convolutional Neural Network (CNN) and Residual Network for the \textit{Deep Learning Architectures} category and Normal Distribution for the \textit{Probability and Statistics} category. We also notice that the identified DL framework pairs associated with each category reveal cross-framework code cloning instances, and 8 out of the nine studied DL frameworks present at least one cross-framework file-level clone. \\

\noindent\textbf{The computed lifetimes of cross-framework file-level pairs vary across categories, ranging from 0 to 8 quarters, i.e., quarters.} We compute the lifetimes of cross-framework file-level clone pairs to reflect how long a particular clone pair persists within a specific functional category over time. As shown in table \ref{tab:functional_code_clone_distribution}, some categories, such as \textit{Communication Interface}, experience short-lived cloning instances lasting less than a quarter, while other categories, such as \textit{Version Control} and \textit{Probability and Statistics} persist over a more extended period between 5 and 8 quarters. This variation suggests that code within different functional categories evolves at different rates. Some categories, such as \textit{Version Control} and \textit{Probability and Statistics}, may involve functionalities with more stable and fundamental design patterns less prone to frequent changes, resulting in longer lifetimes. On the other hand, categories like \textit{Communication Interface} may experience frequent updates or changes, leading to shorter lifetimes. These insights highlight areas where stability or adaptability is crucial when managing and maintaining code reuse.\\

\begin{table}
\centering
\caption{The distribution of across framework functional code clones.}
\label{tab:functional_code_clone_distribution}
\small
\begin{tabular}{lccl*{4}c}
\toprule
      \textbf{\makecell{Functional code clone\\categories}} & \textbf{\makecell{\# of file\\clone pairs}} & \textbf{\makecell{Clone pair \\lifetime range}} &
      \textbf{\makecell{DL framework\\pairs}}
      \\
   \midrule

Communication Interface & 1 & 0 quarter& (\textit{BentoML}, \textit{Paddle}) \\
Distributed Training in TensorFlow & 1& 2 quarters& (\textit{Ray}, \textit{TensorFlow})\\
Python Object Serialization & 1& 1 quarter& (\textit{Aesara}, \textit{BentoML})\\
Efficiency in Python & 2& 3 to 4 quarters & (\textit{BentoML}, \textit{TensorFlow}), (\textit{Paddle}, \textit{PyTorch})\\
Version Control & 2& 8 quarters& (\textit{Ray}, \textit{BentoML})\\
Deep Learning Architectures & 5& 0 to 2 quarters& (\textit{Ray}, \textit{TensorFlow}), (\textit{Ray}, \textit{PyTorch})\\
Probability and Statistics & 8& 5 quarters& (\textit{JAX}, \textit{TensorFlow}), (\textit{MXNet}, \textit{PyTorch})\\ \midrule
Total & 20 & \\ 
\bottomrule
\end{tabular}
\end{table}

\noindent\textbf{Functional code clones across DL frameworks undergo a gradual disappearance, attributed to functionality evolution, code divergence, function deprecation and framework restructuring.} Over time, code functionality evolves and leads to a divergence from the initial commonality. For instance, an initial commonality, such as a normal distribution, may undergo changes, as observed in \textit{Jax}, where the file evolves to encompass additional functionalities like special functions, such as the gamma function, leading to a disappearance of the code clone across frameworks. Code divergence is another reason often resulting from adopting third-party libraries. For example, TensorFlow's shift to a new third-party library for object serialization in Python diverges from the code used in \textit{BentoML}, which continues to use Cloudpickle. Additionally, function deprecation also leads to the decline of functional clones as frameworks undergo updates and discontinue specific functionalities. Furthermore, framework restructuring and the creation of new separate modules lead to the removal of certain functionalities. For example, the removal of Residual Network (ResNet) implementation to a standalone module outside of \textit{PyTorch} led to the disappearance of clones between \textit{PyTorch} and \textit{Ray}.\\

\noindent\textbf{\textit{Architectural adaptation clones} between Keras and \textit{TensorFlow} are performed gradually over multiple releases.} \textit{TensorFlow} and \textit{Keras} are the only two DL frameworks among the studied ones that present \textit{architectural adaptation clones}. The evolution of cloned files between these two frameworks suggests a dynamic and evolving integration process with periods of stability, optimization, and significant updates. For instance, between March 2018 and December 2019, we observe a relatively stable number of file clone pairs, ranging from 73 to 42 pairs. This suggests an ongoing integration process where \textit{TensorFlow} and \textit{Keras} modules are aligning and adapting. For instance, in release  \textit{TensorFlow v1.13.1}\footnote{https://github.com/tensorflow/tensorflow/releases/tag/v1.13.1}, a new functionality analogous to "tf.register\_tensor\_conversion\_function" was added. Then, through the module adaptation process, these two frameworks present a decrease in the cloned file that is due to significant refactoring efforts within the \textit{Keras} library, such as the exclusion of the two modules "applications" and "preprocessing" from Keras as described in the release \textit{Keras 2.2.0}\footnote{https://github.com/keras-team/keras/releases/tag/2.2.0} notes. At a later stage, \textit{TensorFlow} and \textit{Keras} present a very small number of clones (i.e., between 0 and 5 pairs). The decrease in cloned code file pairs is attributed to optimization efforts within \textit{TensorFlow} focusing on refining and optimizing the previously integrated \textit{Keras} code while the development is discontinued in \textit{Keras}, marking the full transition to the \textit{TensorFlow} codebase. \\

\begin{Summary}{}{firstsummary} DL frameworks present \textit{functional} and \textit{architectural adaptation} code clones. \textit{Functional code clones} in DL frameworks span seven categories and gradually disappear due to functionality evolution, code divergence, function deprecation, and framework restructuring. The most frequent categories, \textit{Probability and Statistics} and \textit{DL Architectures}, suggest commonalities in key features across different DL frameworks. This implies opportunities for collaboration, code reuse, and understanding best practices in DL framework development. Architectural adaptation clones represent the integration of one framework module into another, such as the integration of \textit{Keras} into \textit{TensorFlow}. \end{Summary}

\section{Implications}
\label{sec:discussion}
In this section, we discuss the possible implications of our findings that could be useful to practitioners and researchers.\\

\noindent\textbf{Fostering collaboration in clone maintenance.} In RQ2, our results show that the number of committers to a clone pair is a significant factor influencing cloned code size increase, as indicated by the highest \(\chi^2\) in $Model_{Ascending}$, implies that minimizing the number of developers involved in modifying a clone pair may contribute to more consistent and well-maintained code clones. This suggests that a smaller team with a better understanding of the codebase and, more specifically, the cloned code is more effective in propagating changes consistently, thus reducing the likelihood of inconsistent copies. Researchers can leverage these insights to explore strategies for code clone management and team collaboration. In fact, the observation in RQ3 that there exists a consistently low community involvement in code clones complements the findings of RQ2 and underscores the need to encourage broader community engagement in clone-related activities. Project maintainers and open-source community leaders should proactively foster collaboration, knowledge-sharing, and contribution initiatives in clone maintenance. For example, enhancing the documentation can promote a wider understanding of clone-related tasks and attract diverse contributors \cite{wong2019improving}, mitigating the reliance on original authors and enhancing the sustainability and robustness of DL framework development communities.\\

\noindent\textbf{Simplifying abstract class complexity}. Addressing code complexity in abstract classes is crucial to mitigating code clone proliferation, as complexity emerges as the second most explanatory factor in $Model_{Ascending}$ and $Model_{Descending}$. By emphasizing code modularization and reuse strategies, developers can potentially reduce the need for code cloning and enhance overall code maintainability. This insight underscores the importance of efficient coding practices and architectural design for DL frameworks to minimize the challenges associated with class structures. For instance, previous studies \cite{Morshed, 1011328} demonstrate that code clones can lead to a lack of good inheritance structure or abstraction. \\

\noindent\textbf{Optimizing code maintainability through effective maintenance strategies}. As demonstrated in $Model_{Stable}$, the stable cloned code size is associated with effective maintenance strategy in code clones, such as the reduction of the count of declaration statements and the promotion of a more concise and modular structure by using abstract classes. Hence, developers should prioritize the creation of smaller, modular code segments to enhance maintainability and ease of understanding of clone code.\\

\noindent\textbf{Addressing quality challenges in \textit{"thick clones"}}. In RQ1, we observe that a significant number of bugs, i.e., more than 50\% of bugs for each DL framework, are associated with \textit{"thick clones"}. This observation emphasizes the potential risks posed by larger and more complex clones, which highlights the importance of thorough code reviews, testing, and bug tracking in sections of code characterized by high clone thickness. Prioritizing bug detection and resolution in \textit{"thick clones"} is crucial for maintaining software quality and reliability.\\

\noindent\textbf{Standardizing code practices with DL frameworks.} The results of RQ3 demonstrate that some cross-framework clones are prevalent in specific functional codes, namely \textit{Probability and Statistics} and \textit{Deep Learning Architectures}. These results imply a potential for standardized practices or shared code components in some DL common functionalities, such as architectural design, offering opportunities for collaboration, code reuse, and a deeper understanding of best practices in DL framework development.\\

\section{Threats to Validity}
\label{sec:threats}
\noindent\textbf{Construct validity} relates to a possible error in the data preparation. There is no established tool available to identify all existing DL frameworks. In our study, we manually curate DL frameworks from GitHub which could lead to errors. To address this potential concern, we conduct cross-validation on our selected frameworks with the most recent studies  \cite{10.1145/3338906.3338955, 10.1007/s10664-021-10099-x, 10.1145/3368089.3409761, 10.1145/3587155, Humbatova_202, 8355444, 10.1145/3514094.3534167, 9463133, Wang2019} on DL frameworks. Our 
study stands as one of the most extensive DL framework comparative investigations to date, encompassing a substantial number of Python-based DL frameworks, i.e., nine frameworks, in a single comprehensive analysis, which greatly enhances its contribution to the understanding of this field. Therefore, we assert a strong construct validity. 

Another potential threat could be related to the choice of the clone detector in our study. We employed NiCad to identify code clones within each DL framework snapshot, as it has been a common choice in previous research \cite{10.1007/s10664-021-10099-x, 10.1145/3607181}. However, we acknowledge that NiCad's configuration settings can influence the outcomes of our clone detection process. To address this concern, we tailored the configuration settings based on the methodologies outlined in prior studies \cite{10.1007/s10664-021-10099-x, 10.1145/3607181}, which have demonstrated to achieve high precision and recall rates.

Lastly, we adopt the heuristic introduced by Barbour et al. \cite{883028} to investigate the code clone bug proneness to identify bug-fix commits. This involves utilizing a predefined set of keywords associated with bug fixes. If any of these keywords are found within a commit's message, it is categorized as a bug-fix commit. This heuristic might lead to inaccurately classified commits. However, it has been successfully used in multiple previous studies from the literature \cite{8919065, 10.1007/s10664-021-10099-x, 10.1145/3607181, 6080794}, and proven to achieve satisfyingly accurate results.\\

\noindent\textbf{Internal validity} relates to the concerns that might come from the internal methods used in our study. In RQ2, we offer explanations for these observed results related to practitioner development practices. However, we refrain from asserting causation and instead provide observations and correlations.\\

\noindent\textbf{External validity} relates to the potential of generalizing our study results. Given that most DL frameworks are developed in Python, our study primarily concentrates on frameworks implemented in Python. Consequently, we cannot ensure the broad applicability of our findings to frameworks developed in different programming languages, such as C++ and Java. However, we have made an effort to encompass a wide range of DL frameworks of different sizes and functionalities.

\section{Related Work}
\label{sec:related_work}
In this section, we present a review of the existing literature on analyzing DL frameworks. Additionally, we touch upon the studies relevant to code clone research.

\subsection{DL framework-related empirical studies}

DL, emerging as a prominent field in recent years, has attracted considerable attention within the software engineering community. A growing number of researchers have been actively empirically investigating the characteristics of DL applications and frameworks. While some researchers focused their interest on DL applications, such as identifying the challenges of building DL-based systems \cite{8987482, morovati2023common} and exploring the taxonomies of faults in DL applications \cite{Humbatova_202, 10.1145/3379597.3387479}, others direct their attention toward the exploration of the DL frameworks that are the building blocks of these applications.\\

\noindent\textbf{Bug taxonomies and characteristics.} Researchers contribute to understanding DL framework bugs comprehensively in several ways. TensorFlow, one of the most popular DL frameworks, has gained the attention of several researchers \cite{JIA2021110935, 10.1145/3213846.3213866}. For instance, Zhang et al. \cite{10.1145/3213846.3213866} analyze TensorFlow bugs sourced from StackOverflow QA pages and GitHub. The authors provide taxonomies along several dimensions, including root causes, symptoms, and bug detection strategies. Other researchers expand their work to investigate multiple DL frameworks. Chen et al. \cite{10.1145/3587155} conduct a comprehensive analysis of four frameworks, i.e., TensorFlow, PyTorch, MXNet, and DL4J, to identify root causes and symptoms and provide actionable guidelines for improved bug detection and debugging. In addition, the authors develop a tool, called TenFuzz, to identify bugs in Tensorflow. Similarly, Islam et al. \cite{10.1145/3338906.3338955} analyze the characteristics of bugs in five DL frameworks. In addition, the authors identify the stages of the DL pipeline that are more bug-prone and investigate antipatterns. Tambon et al. \cite{Tambon2021SilentBI} sheds light on silent bugs in DL frameworks. The authors classify the bugs according to their impact on users' programs and the specific components where these issues originated, drawing upon information found in the issue reports. Du et al. \cite{9716780} provide a classification of the fault-triggering conditions of bugs.
The authors manually investigate 3,555 bug reports collected from three TensorFlow, MXNET and PaddlePaddle and analyze the frequency distribution of different bug types and their evolution features. Long et al. \cite{10.1145/3530019.3530029} present an exploratory study on performance and accuracy bugs in ten popular DL frameworks, revealing insights such as the primary root cause for reporting performance bugs, and they offered actionable implications for researchers, maintainers, and submitters to improve the bug management process of performance bugs.\\

\noindent\textbf{Bug fixing patterns.} Jia et al. \cite{JIA2021110935} explore TensorFlow, offering insights into the bugs and their fixing process within the framework's components. Ho et al. \cite{ho2023empirical} replicate the study by Jia et al. to explore another popular DL framework, PyTorch. In addition to identifying the bug-fixing patterns in PyTorch, the authors provide a detailed comparison between TensorFlow and PyTorch, highlighting the similarities and differences between the frameworks' bugs. Islam et al. \cite{10.1145/3377811.3380378} conducted a larger scope study encompassing five DL frameworks. The authors investigate the challenges associated with 970 bug-fix patterns from Stack Overflow and GitHub and find that the most common patterns are related to data dimension and neural network connectivity issues. Li et al. \cite{li2023understanding} follow a different direction by focusing on multi-language DL frameworks. Apart from exploring the bug types and their impacts on DLF development, the authors discover that addressing bugs in multi-language frameworks involves significantly greater complexity in code changes compared to single-programming-language bug fixes.\\

\noindent\textbf{Technical debt.} 
Liu et al. \cite{10.1007/s10664-020-09917-5} investigate the DL framework from a technical debt perspective. The authors analyze the comments indicating technical debt(self-admitted technical debt) of seven popular DL frameworks and identify seven types of technical debt in DL frameworks, i.e., design, defect, documentation, requirement, test, compatibility, and algorithm debt.\\

While the above work also focuses on studying DL frameworks, it only covers aspects related to bug taxonomies, bug-fixing patterns and technical debt. Our work offers an orthogonal study that investigates the evolution of clones and their impact on the maintenance of DL frameworks alongside bug-related aspects.

\subsection{Code Clones}

\subsubsection{Code clones analysis in traditional systems}
Researchers have extensively explored code clones in traditional systems \cite{10.1145/3569966.3570091}. These studies encompass a multifaceted exploration of code clones, including their impact on software maintenance, bug-proneness, change-proneness, and evolution.\\

\noindent\textbf{Impact on software maintenance.} Existing work demonstrates that code cloning could result in increased maintenance challenges for software systems \cite{7880507, 8327317, 4145027, 8327315}. Mondal et al. \cite{7880507} conduct a comparative empirical study to investigate the maintenance efforts required for cloned and non-cloned functions. The study found that cloned code is associated with an increased maintenance cost, in particular with Type 2 and Type 3 clones. Higo et al. \cite{8327317} demonstrate that among web-based systems developed from the same specifications, those projects with a high prevalence of code clones present a greater challenge for project testing. In particular, the study demonstrates an increased effort in bug detection during unit testing and an increased number of clones during integration testing. Islam et al. \cite{8327315} conduct a comparative analysis of code clones with and without bugs, considering 29 code quality metrics across 2,077 revisions of three Java software systems and offering insights for cost-effective clone management and clone-aware software development.\\

\noindent\textbf{Bug proneness.} Several previous studies have examined the bug proneness of code clones, revealing that clones make code more bug-prone and increase maintenance costs \cite{8090146, Islam2017ACS, 8667993, 7476632, 10.1145/1287624.1287634}. Islam et al. \cite{Islam2017ACS} compare the bug-proneness of clone and non-clone code. This paper presents a comparative study on the bug proneness of code clones and non-clones, finding that clone code has a significantly higher percentage of files changed due to bug-fix commits and a greater likelihood of severe bugs, suggesting that bug-fixing changes in clone code require extra attention. In another study, Islam et al. \cite{8667993} delve into the bug-proneness of micro-clones, i.e., small code fragments of 1 to 4 lines of code and compare them with regular code clones in diverse open-source systems written in C, C\#, and Java. The study findings show that micro-clones are significantly more prone to bugs, exhibit more consistent changes due to bug-fix commits, affect a higher percentage of files, and contain a greater percentage of severe bugs than regular clones, hence underscoring the importance of managing and maintaining micro-clones in software development. Jiang et al. \cite{10.1145/1287624.1287634} study also validates that code clones had a higher likelihood of containing bugs, primarily originating from inconsistencies within cloned code segments.\\

\noindent\textbf{Change proneness.} The negative impacts of code clones on software maintenance have led to extensive research on clone stability and change proneness. \cite{8665850, 10.1007/s10664-017-9528-y, Mondal2021, 10.1145/1808901.1808910, 4658071, mondal2020fine}. 
For example, Mostafa \cite{8665850} focuses on analyzing clone evolution with respect to clone location, i.e., Inter-File and Intra-File, and clone lifetime. The study reveals that Intra-File clones are more prevalent, suggesting that developers tend to duplicate code within the same file, and these clones are also more dynamic, indicating a preference for refactoring or altering clones within the same file. Mondal et al. \cite{10.1007/s10664-017-9528-y} find that cloned code tends to be more change-prone and unstable during the maintenance phase than non-cloned code. The study also identifies differences in stability among various types of clones, programming languages, and programming paradigms. In a more recent study, Mondal et al. \cite{Mondal2021} conduct a comprehensive study on 12 subject systems to compare the stability of clone and non-clone codes. The authors demonstrate that code clones generally exhibit higher change-proneness as compared to non-clones by referring to the eight stability measuring metrics, implying increased maintenance effort and cost.\\

\noindent\textbf{Clone evolution.} Some researchers have constructed clone genealogies by tracing the history of code clones \cite{10123634, barbour2018investigation, 10.1007/s10664-018-9645-2}. Barbour et al .\cite{barbour2018investigation} derive six distinct evolutionary patterns by building the clone genealogies of four open-source Java systems. In addition, the authors leverage the clone genealogy information to enhance the effectiveness of fault prediction models. Similarly, Thongtanunam et al. \cite{10.1007/s10664-018-9645-2} conduct an empirical study on six open-source Java systems genealogies, revealing that a significant proportion of clones have short lifespans. In addition, the authors predict, using random forest classifiers, whether a newly introduced clone would be short-lived based on factors extracted from the genealogy. In a recent work, Bladel and Demeyer \cite{10123634} conduct a study on eight open-source systems genealogies and revealed that code clones are more prevalent in test code compared to production code due to the recurring pattern of unit test code.\\

The studies mentioned above highlight the challenges of code cloning in traditional software on software maintenance, including bug proneness, change proneness and the clone evolution. However, given the differences between traditional and DL frameworks, we cannot assume those findings directly apply. Our work aims to address the knowledge gap concerning the evolution of code cloning and its implications within the DL frameworks.

\subsubsection{Code clones analysis in DL systems}

Despite the increasing surge in the development of DL software, only two studies have investigated code clones within this domain, highlighting the need for more comprehensive investigations into code cloning practices in DL software. 

Jebnoun et al. \cite{10.1007/s10664-021-10099-x} introduce the first study on clones in DL development. The authors analyze code clones in 59 Python, 14 C\#, and 6 Java-based DL systems and an equal number of traditional software, highlighting the frequency, distribution, and effects of code clones. In addition, they provide a taxonomy to identify phases with a higher risk of cloning-related faults. Their findings indicate that code cloning is prevalent in DL systems, with developers often copying code from files located in other directories, and that code cloning is more common during DL model creation, training, and data preprocessing phases. In addition, the authors find that code clones in DL code are likely to be more defect-prone compared to non-cloned code. 

Similarly, Mo et al. \cite{10.1145/3607181} also find that code clones are prevalent in DL systems, exhibiting nearly twice the rate in traditional projects. However, the authors conduct their study on a dataset of Python projects with only 45 DL and 45 traditional projects. Different from the work of Jebnoun et al., Mo et al. focus on co-changed clones and investigate the distribution of Type 1, Type 2, and Type 3 co-changes and their bug-proneness. The authors also conduct a comparative analysis of the DL applications subject based on the underlying DL frameworks.

This paper conducts an empirical study of code clone evolution in DL frameworks. Our work differs from the aforementioned previous work in two ways: firstly, we study code clones in DL frameworks rather than the DL applications, and secondly, our analysis focuses on the evolutionary and comparative aspects of code clones as opposed to merely studying the distribution of clones within a single snapshot. Moreover, we investigate the code clones across DL frameworks and not only within a framework. We also examine the clone trend within releases and provide developers with insights for stable, maintainable clone practices.

\section{Conclusion}
\label{sec:conclusion}
Given the pivotal role of DL frameworks in the rapidly growing field of artificial intelligence and machine learning, its software quality is crucial for the development of DL-based applications. Code clone is a common coding practice that can potentially adversely affect software maintainability. In this paper, we conduct an empirical analysis of nine popular DL frameworks, including TensorFlow, Paddle, PyTorch, Aesara, Ray, MXNet, Keras, Jax, and BentoML, to reveal insights into the long-term code clone trend over releases, and within-release clone patterns, i.e., short-term, and their implications. In addition, we also conduct a file-level cross-framework clones analysis to identify the cross-functionalities of cloned code within the frameworks.

Our results identify four distinct trends in code clone evolution. In addition, we provide an understanding of the factors influencing cloned code size reduction and unveil the association between within-release clone patterns and long-term clone trends. Our research contributes to the foundational understanding of code cloning in the context of DL frameworks. The identification of cross-framework file-level code clones, categorized as \textit{functional} and \textit{architectural adaptation} code clones, highlights collaborative opportunities and commonalities in key features across different DL frameworks. Moreover, our study offers insights emphasizing the importance of fostering collaboration, simplifying abstract class complexity, optimizing code maintainability, and standardizing code practices within DL frameworks. As the DL landscape continues to evolve, our findings provide valuable insights to the DL community, contributing to the sustainable development and maintenance of these critical systems. 

In our future work, we plan to broaden the scope of our study beyond Python projects to include other programming languages, such as C++ and Java, aiming for a more comprehensive understanding of code cloning in diverse DL framework language ecosystems.

\begin{acks}
The first author would like to acknowledge the support of NSERC Vanier Canada Graduate Scholarships. 
\end{acks}

\bibliographystyle{ACM-Reference-Format}
\bibliography{sample-base}

\appendix

\end{document}